# Topological Spatiotemporal Mode-Locked Lasers

Xiaokang Lian[1,2,3], Yuchen Zhang[1,2,3], Mutian Xu[1,2,3], Bing Han[1,2,3], Chengkun Cai[1,2,3], Jun Liu[1,2,3], Jian Wang[1,2,3*]

[1]Wuhan National Laboratory for Optoelectronics and School of Optical and Electronic Information, Huazhong University of Science and Technology, Wuhan 430074, Hubei, China

[2]Hubei Optical Fundamental Research Center, Wuhan 430074, Hubei, China

[3]Optics Valley Laboratory, Wuhan 430074, Hubei, China

* Corresponding author: Email: jwang@hust.edu.cn

**Abstract:** Achieving robust high-dimensional solitons both in optics and Bose-Einstein condensates remains a great challenge, due to unpredictable high-order perturbations. Here we propose and demonstrate a novel topological spatiotemporal mode-locked fiber laser empowered by mode-division multiplexing, for precisely controlling of the spatiotemporal solitons. By introducing time delays between the individual transverse mode paths, a topological spatiotemporal lattice is formed in the multimode cavity, which in turn excites periodic spatiotemporal solitons. These solitons are stable against the perturbations, attributed to periodic potentials induced by the nonreciprocal dissipative coupling interactions. Moreover, under the thermodynamic limit, the solitons condense into nested-type spatiotemporal solitons, a phenomenon analogous to the Bose-Einstein condensate of matter waves. Our results indicate that topological spatiotemporal mode-locked lasers not only constitute a controllable source of robust spatiotemporal light bullets, but also provide a valuable platform for exploring nonlinear high-dimensional soliton dynamics and non-Hermitian topological physics.

## Introduction

Solitons are coherent, localized wave states that arise in nonlinear systems, including Bose–Einstein condensates (BECs), classical and non-classical flows of superfluids, exciton–polaritons, plasmas, quantum gases, waveguides, nonlinear lattices, lasers, and oscillators [1–14]. Compared with their relatively simple one- and two-dimensional counterparts, high-dimensional solitons exhibit substantially richer structures and underlying physics. The generation and manipulation of such high-dimensional solitons are therefore of central importance for a broad range of applications, including high-bandwidth optical communications, quantum information processing and computing, precision quantum metrology, high-power and ultrafast laser sources, material processing, light–matter interactions, optical trapping and micromanipulation, as well as fundamental atomic and molecular physics [15–18]. Despite decades of intensive theoretical and experimental efforts, however, the realization of coherent high-dimensional solitons remains challenging, and experimental demonstrations have been limited to only a few platforms.

Nonlinear multimode systems provide a powerful platform for exploring high-dimensional optical solitons [19–22]. In multimode fibers (MMFs), however, the optical field is typically highly complex and often chaotic, owing to multimode interference and group-velocity walk-off among hundreds to thousands of transverse modes, further exacerbated by fabrication-induced imperfections and environmental perturbations. Recent experiments have demonstrated that spatiotemporal solitons can nevertheless emerge from a delicate balance between intramodal dispersion and multimode nonlinear interactions [23–25]. Beyond propagating solitons, laser cavity solitons are particularly appealing, as they offer greater flexibility in control and manipulation through cavity engineering [26–28]. It has been shown that multimode mode-locked fiber lasers can generate and tailor coherent spatiotemporal solitons in practical systems [29–37]. These coherent states arise when transverse and longitudinal modes become jointly phase-locked through intracavity processes, including spectral and spatial filtering, gain–loss balancing, and the interplay between intermodal and intramodal dispersion with multimode spatiotemporal nonlinearities.

Mode-locking in multimode fiber lasers arises from the collective dynamics of many interacting modes, and achieving precise control over the resulting complex spatiotemporal soliton field remains a significant challenge. One obstacle is the presence of multiple attractors in the multimode cavity, which makes the mode-locked state highly sensitive to environmental perturbations and cavity-parameter variations. A second obstacle stems from the thermodynamic behavior of multimode field, exemplified by the multimode self-cleaning effect, in which optical power is transferred from higher-order modes to the fundamental mode under strong nonlinearity [21,22]. Although self-cleaning can improve beam quality [31], it may suppress the formation of highly multimode spatiotemporal solitons [29] and can impede the engineering of more complex spatiotemporal beam structures.

In this work, we apply mode-division multiplexing (MDM)— a technique commonly used in fiber-optic communications [38]—to multimode mode-locked fiber lasers, enabling precise control of high-dimensional spatiotemporal wave packets. The MDM-based lasers feature a composite multi-branch cavity, analogous to coherent pulse stacking and beam combining in single-mode laser arrays [39,40]. We further incorporate a delay-induced non-Hermitian topological protection mechanism [41-48]. The topological architecture introduces an effective periodic potential within the multimode cavity, enabling confinement and stabilization of spatiotemporal solitons. This mechanism is fundamentally distinct from stabilization based on conventional cavity effects—such as spectral and spatial filtering, spatially dependent gain, or saturable absorption—employed in previous single-cavity multimode fiber lasers [29-37]. The topological origin of the confinement enhances robustness against structural perturbations, reduces sensitivity to initial conditions, and enables scalable and reconfigurable system architectures.

## Configuration and principle

Figure 1a shows the configuration of topological spatiotemporal mode-locked fiber laser. The proposed laser system is formed by a multimode fiber (MMF) and an MDM module. The MDM module comprises a mode demultiplexer (DEMUX) and a mode multiplexer (MUX), interconnected by single-mode fiber (SMF) branches. This configuration allows for independent control of the transverse modes in the MMF system. Figure 1a depicts the laser setup employing a six-branch MDM configuration. In this system, six transverse modes in orbital angular momentum (OAM) mode basis of the MMF are separated by the DEMUX and respectively routed into six single-mode branches. After propagation, they are recombined by the MUX back into their original transverse modes. Each single-mode branch is equipped with a pump source, gain fiber, saturable absorber, polarization controllers, and the optical time delay lines (OTDLs).

Remarkably, precise adjustment of the OTDLs can induce a topological structure within the cavity and effectively enhance the mode-locking stability of the MMF laser. As a representative, this is demonstrated below using a laser with a two-branch configuration. As illustrated in Fig. 1b, the two-branch configuration consists of two SMF branches between points A and B. The upper branch (branch-1, blue in color) is assumed to be corresponding to the fundamental mode (mode 1), while the lower branch (branch-2, green in color) is corresponding to a high-order mode (mode 2). The former is longer than the latter by a time delay of $\Delta t$. By narrowing down the common section (orange) until points A and B coincide, the configuration in Fig. 1b can be topologically transformed into the coupled laser oscillators shown in Fig. 1c. Since passively mode-locked fiber lasers typically operate in fundamental or high-order harmonic states due to longitudinal mode locking, it can be inferred that, when the upper oscillator operates at the $(N+1)^{\text{th}}$ harmonic, the lower oscillator operates at the $N^{\text{th}}$ harmonic (Supplementary Materials Section 2). As illustrated in Fig. 1d, the two coupled oscillators are synchronized in frequency, and the time delay between adjacent pulses in either upper or lower ring remains equal to $\Delta t$, controlled by the OTDL. Here, $u_n^m$ and $v_n^m$ denote the optical field at site $n$ in the upper and lower oscillators, respectively, during the $m^{\text{th}}$ round trip.

If the upper oscillator is designated as the main cavity, the lower oscillator can be regarded as a delay-line reservoir [48], where the function of gain and saturable absorber is considered to maintain the soliton shape. This reservoir stores a portion of light from the main cavity via crosstalk, delays it, and then returns it back to the main cavity. In this framework, the pulse lattice is mathematically described by the equations provided in Supplementary Fig. 4b as follows:

$$\begin{cases} u_n^{m+1} = \sqrt{1-\rho}\,u_n^m + \sqrt{\rho}\,v_{n+1}^m \\ v_n^{m+1} = \sqrt{\rho}\,u_n^m + \sqrt{1-\rho}\,v_{n+1}^m \end{cases}. \tag{1}$$

It can be inferred that $u_n$ remains stationary at its lattice position as indicated by the first equation of equation (1) during the periodic process, while $v_n$ moves forward stepwise to a smaller $n$ as shown in the second equation of equation (1). The dynamic pulse $v_{n+1}$ transfers part of its power in the form of $\sqrt{\rho}v_{n+1}^m$ to $u_n$, with a directional (i.e., nonreciprocal) dissipative coupling efficiency of $\sqrt{\rho}$ (Supplementary Materials Section 2.1). This resembles a unidirectional Hatano-Nelson lattice with an imaginary vector potential [43-45], as represented by the broad green arrow in the upper oscillator in Fig. 1d. The reconfigured periodic photonic lattice of $u_n$ derived from Fig. 1d is illustrated in Fig. 1e, where the nonreciprocal dissipative coupling is indicated by green arrows. Similarly, if the lower oscillator is set as the main cavity and the upper as the reservoir (Supplementary Fig. 4c), the distribution of $v_n$ would also form a unidirectional Hatano-Nelson lattice, but with an imaginary vector potential from smaller to larger lattice indices, as indicated by the broad blue arrow in the lower oscillator in Fig. 1d.

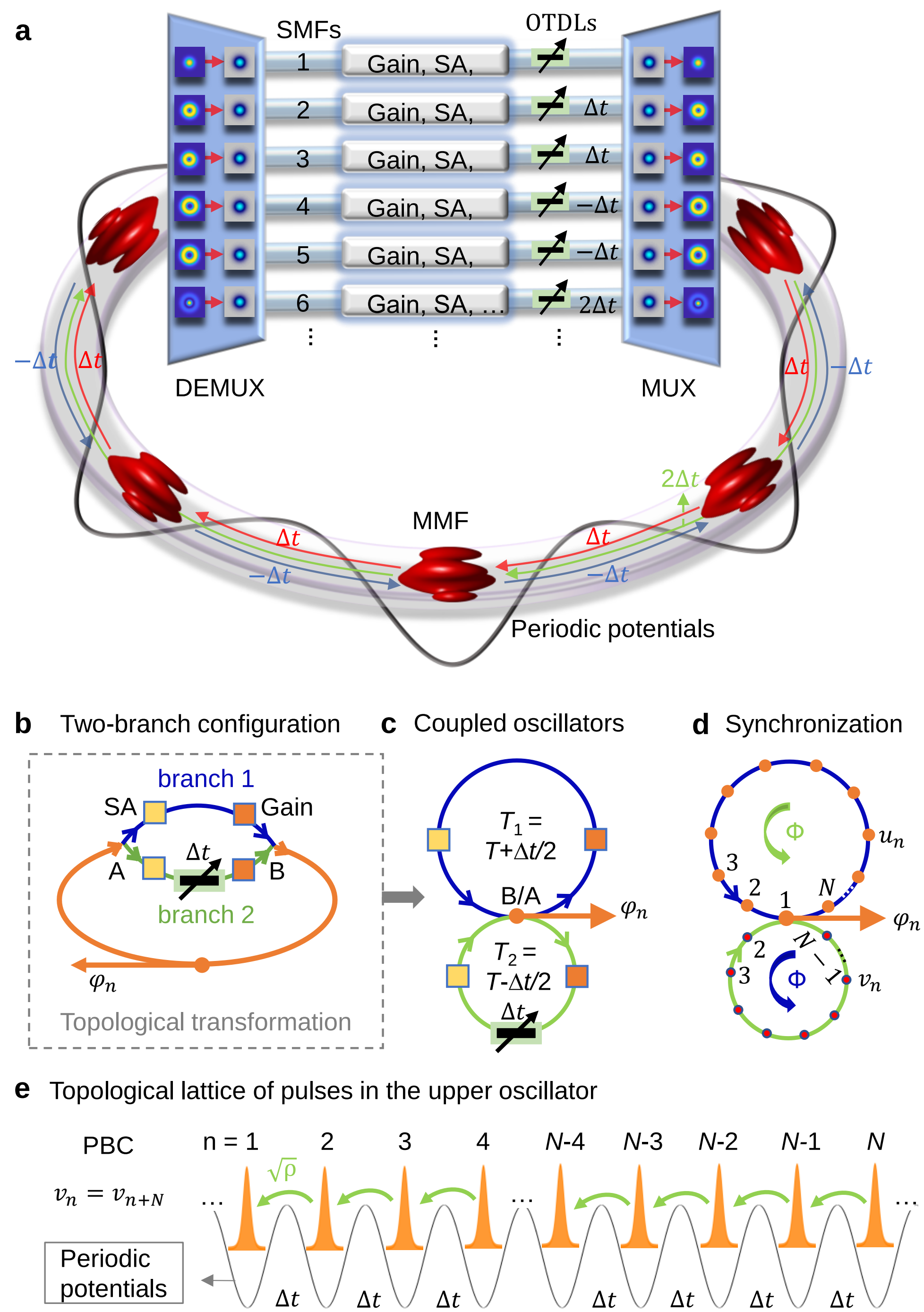

**Fig. 1| Concept and principle of the topological spatiotemporal mode-locked fiber laser.** **a**, Schematic diagram of the topological spatiotemporal mode-locked laser based on the mode-division multiplexing (MDM) configuration. MMF: multimode fiber; SMF: single-mode fiber; MUX: multiplexer; DEMUX: demultiplexer; SA: saturable absorber; OTDL: optical time delay line. **b**, A representative two-branch implementation of the concept. **c**, Topological transformation of the configuration in **b** to a model of coupled laser oscillators. **d**, Illustration of pulse synchronization and mode-locking dynamics in the coupled oscillators. **e**, The photonic lattice with nonreciprocal dissipative coupling.

The dynamics of the two-branch spatiotemporal mode-locked fiber laser can be understood with the periodic photonic lattice shown in Fig. 1e. The governing equation can be formulated as following (Supplementary Materials Section 2.2):

$$\frac{\partial\psi(T,t)}{\partial T} = (\mathcal{D} + \mathcal{G} + \mathcal{N})\psi(T,t) + k\psi(T,t-\Delta t). \quad (2)$$

Equation (2) without the last term is a complex Ginzburg-Landau equation (CGLE) with dispersion ($\mathcal{D}$), gain and loss ($\mathcal{G}$), and the nonlinear saturable absorber and materials ($\mathcal{N}$) [49]. The last term is the intracavity coupling term, resulted from the delay-line reservoir [48]. Where $\psi(T,t-\Delta t)$ represents the field coupled through the reservoir branch with time delay of $\Delta t$ and '$k$' represents the coupling efficient. The intracavity coupling term bring inhomogeneity (periodic modulation/potential) along the fast time ($t$) coordinate in the mean-field model of CGLE. Therefore, the equation (2) is in a sense analogous to the Gross-Pitaevskii equation (GPE). The GPE is an inhomogeneous nonlinear Schrodinger equation, which is usually employed to describe the Bose-Einstein condensate (BEC) of matter waves [16-18].

A key characteristic of the non-reciprocal dissipative coupling lattice formed in our passively mode-locked laser is its reconfigurability. The pulse period is determined by the time delay between the single-mode fiber branches of the MDM configuration, and the harmonic order can be flexibly turned by the time delay line. The density of the photonic lattice can be increased to very high values as $\Delta t$ approaches zero—a regime difficult to access in actively mode-locked lasers due to the limited modulation bandwidth of intensity modulators [48,49]. As illustrated in Fig. 2a, the harmonic order of the laser scales proportionally with frequency, which is inversely related to the time delay according to $f = 1/\Delta t$. The master model in equation (2) and the tight-binding model (Supplementary Materials Section 2.3) are used to simulate the variation of the pulse lattice with the increase of the density (Fig. 2 and Supplementary Figs. 12-14). Just as the GPE indicates that higher densities of Bose atoms and strong nonlinearity enhance interactions and facilitate BEC, the simulated results verify BEC phenomenon also occurs with the pulses in the studied passively mode-locked laser (Fig. 2b).

As shown in Fig. 2c, continuous-wave like solitons at a harmonic order of $N$ = 492 ($\Delta t = 300$ ps) shrink into envelope solitons when the harmonic order increases to $N$ = 984 ($\Delta t = 75$ ps). Further increasing the harmonic order to $N$ = 7866 ($\Delta t = 9$ ps) compresses the envelope into tightly bound solitons. These results suggest that increasing the pulse density in the laser cavity leads to a form of classical light condensation, analogous to the BEC in matter waves [9,50]. BEC is inherently a nonlinear process, and stronger nonlinearity likewise favors condensation [50,51]. A similar trend is observed in our laser system, i.e., increasing the pump power leads to condensation behavior in simulations. As shown in Fig. 2d, the uniform soliton train (panel 1) collapses into an envelope soliton train (panel 2), which is similar to the transition from panel 1 to panel 2 in Fig. 2c. However, further increasing the pump power does not lead to more pronounced condensation (panel 3). This might be due to the relatively low soliton density—which, as discussed, plays a critical role in facilitating the condensation phenomenon.

When more single-mode fiber branches are incorporated, the spatiotemporal pulse lattice would become more complex, as depicted in Fig. 1a. By setting the delays of the degenerated modes $OAM_{01}$ (mode 2) and $OAM_{0-1}$ (mode 3) in either branch 2 or branch 3 to $\Delta t$, those of $OAM_{02}$ (mode 4) and $OAM_{0-2}$ (mode 5) in branch 4 or branch 5 to $-\Delta t$, and that of mode $OAM_{12}$ (mode 6) in branch 6 to $2\Delta t$, the inter-pulse spacing becomes $\Delta t$, corresponding to a maximum free-spectral range of $1/\Delta t$. Nonreciprocal dissipative coupling is represented by arrows with varying lengths and directions in the MMF section of the laser. The spatiotemporal pulses generated in the cavity are stabilized and confined by periodic potentials resulting from the introduced time delays (Supplementary Section 3). If there are multiple reservoir branches, the last term in the equation (2) can be replaced with $\sum_r^N k_r \psi\,(T, t - \frac{rT_{RT}}{N})$ (Supplementary Section 2.2).

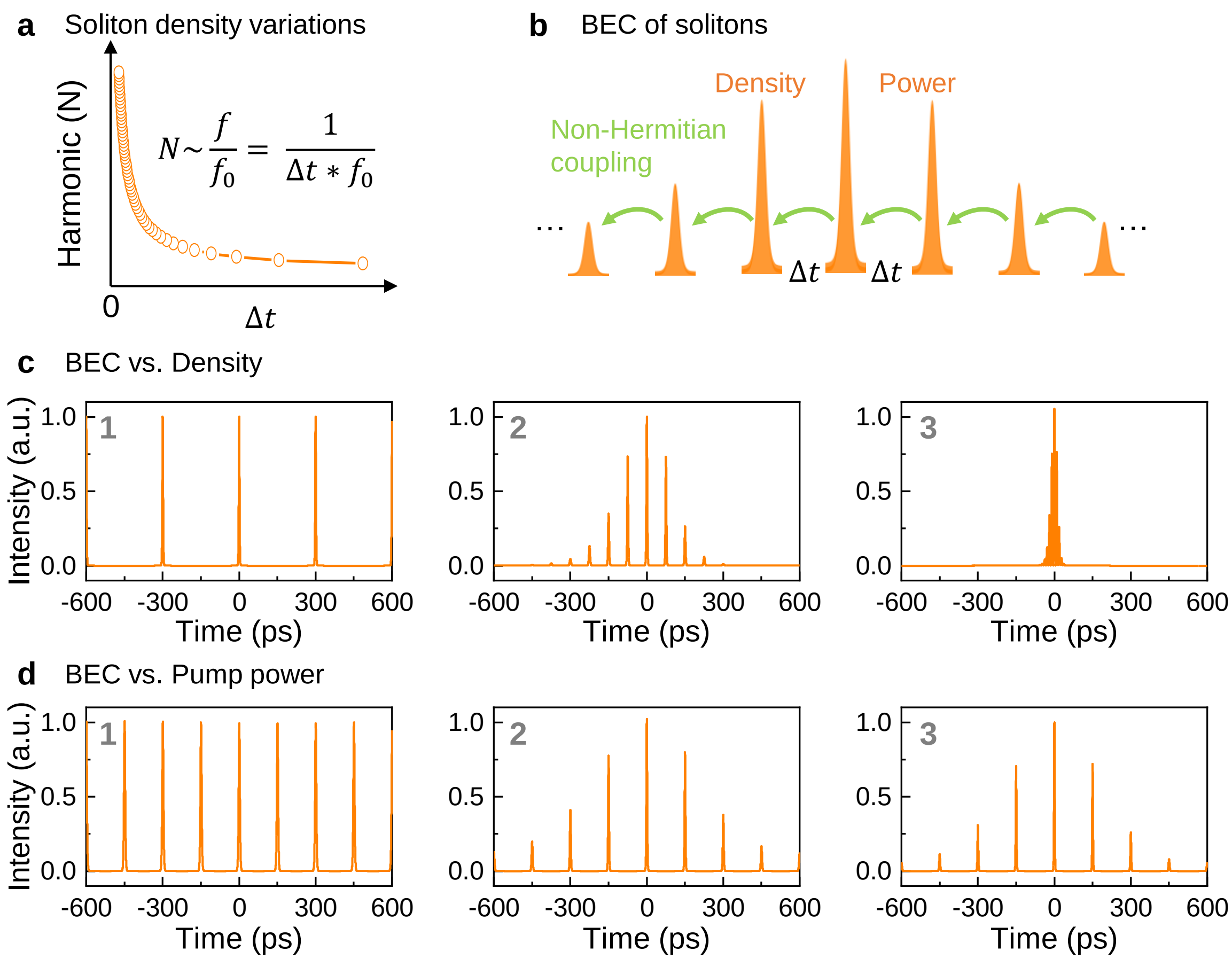

**Fig. 2| Topological lattice soliton condensation in the laser cavity**. **a**, Evolution of the harmonic order of passively mode-locked laser under varying time delay. **b,** Bose-Einstein condensation (BEC) of solitons in the topological spatiotemporal mode-locked laser, governed by soliton density and pump power. **c**, Condensation behavior with increasing soliton density as the time delay $\Delta t$ decreases from 300 ps (panel 1), to 75 ps (panel 2), and to 9 ps (panel 3). **d**, Condensation evolution with increasing pump power as the saturated energy ($E$sat) increases from 300 pJ (panel 1), to 350 pJ (panel 2), and to 400 pJ (panel 3). These results were simulated by equation (2).

## Numerical and experimental verifications

A multimode fiber laser incorporating the two-branch MDM structure is simulated using a lumped element model (Supplementary Materials Section 2.4). The two-branch MDM is specifically designed to control the first two modes, i.e. $OAM_{00}$ and $OAM0_{01}$, in the graded-index MMF. As illustrated in Figs. 3a to 3c, the evolution of mode-resolved spectra, pulse trains, and iso-intensity profiles of spatiotemporal wave packets are presented for different mode-locking states under varying time delays $\Delta t$ as indicated. Despite the presence of ten modes in the simulation results, only the first two modes ($OAM_{00}$ and $OAM_{01}$) contribute dominantly to the power distribution, while higher-order modes remain negligible (panel 4 in Fig. 3a). At a relatively large time delay of $\Delta t$ = 400 ps, the temporal period of the soliton pulse train is 400 ps, as shown in panel 1 of Fig. 3b. The optical spectra in panel 1 of Fig. 3a reveals a modulation period of $\Delta\lambda$ = 0.02 nm, which is consistent with the relation $\Delta\lambda = \lambda^2/(c\Delta t)$ [52]. The inset in panel 1 of Fig. 3c displays a time-integrated beam profile, featuring an eccentric disk with an intensity singularity at its edge. This eccentric disk pattern results from the interference among different transverse modes in the MMF, particularly the first two modes ($OAM_{00}$ and $OAM_{01}$). The iso-intensity profiles of spatiotemporal wave packets (panel 1 in Fig. 3c) exhibit uniform shape and size, corresponding to a regularly spaced soliton pulse train. At such large $\Delta t$ values, the harmonic order remains relatively low, and the soliton density is insufficient to trigger BEC. This regime (group 1 in Fig. 3) is labelled as “Topological”.

For smaller time delays (group 2 with $\Delta t$ = 60 ps and group 3 with $\Delta t$ = 10 ps in Fig. 3), the soliton density becomes high enough to excite BEC. The corresponding pulse trains exhibit intensity modulation envelopes, where pulses at both ends gradually attenuate, as shown in panels 2 and 3 of Fig. 3b. The full width at half maximum (FWHM) of the envelope, fitted with a Sech function, decreases from 400.3 ps to 70 ps as $\Delta t$ is reduced (dashed curves). The spatiotemporal wave packets in panels 2 and 3 of Fig. 3c adopt a rugby ball-like shape that shrinks with decreasing time delay. This type of envelope soliton, comprising two longitudinal frequencies, is referred to as a “nested soliton”, as observed in topological photonic frequency combs [46]. Moreover, the spectral comb tooth in the optical spectra (panels 2 and 3 of Fig. 3a) increases from 0.136 nm to 0.815 nm, reflecting a rise in longitudinal frequency. Given the coexistence of topological and BEC features, groups 2 and 3 are labelled as “Topological + BEC”.

When $\Delta t$ = 0, meaning the lengths of two fiber branches are identical, the delay-induced dissipative coupling and the unidirectional Hatano-Nelson lattice mode are no longer sustained. As a result, the laser enters a topological trivial state and the group 4 is labelled as “Trivial”. The spectrum (panel 4 in Fig. 3a) exhibits a broad bandwidth with Kelly sidebands, a hallmark of soliton mode-locking in the anomalous dispersion regime.

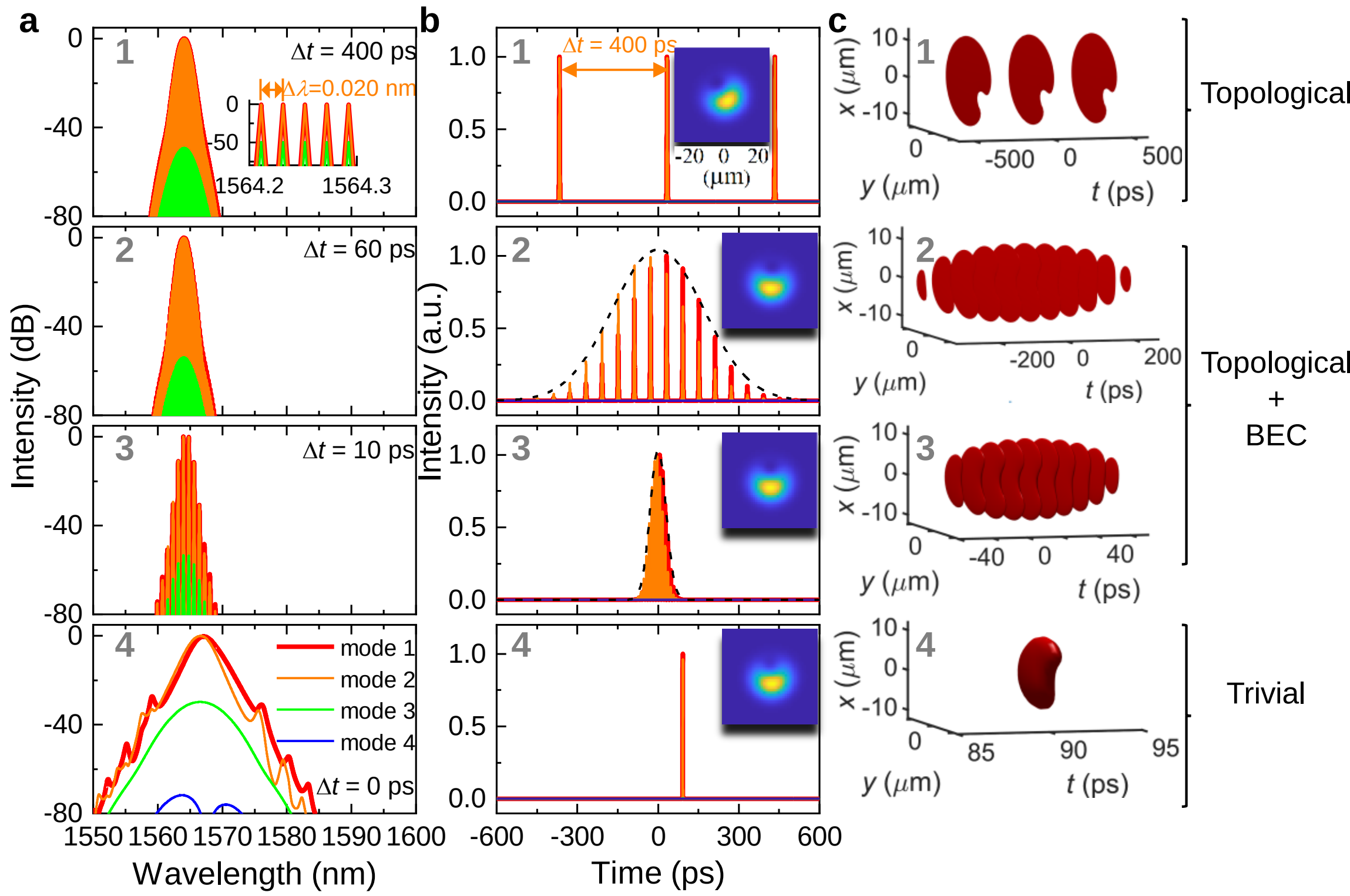

**Fig. 3| Simulated characteristics of the topological spatiotemporal mode-locked fiber laser. a**, Evolution of the laser spectra under different time delays, corresponding to the "Topological", "Topological + BEC", and "Trivial" states. **b**, Corresponding temporal pulse trains and their iso-intensity profiles. **c**, Corresponding spatiotemporal wave packets in iso-intensity profiles. Group 1 to 4 correspond to time delays of 400 ps, 60 ps, 10 ps, and 0 ps, respectively.

The simulated results are further verified by experiments (Supplementary Material Section 1). In the experiments, the repetition period for mode 1 ($OAM_{00}$) of the laser under study is fixed at 73.5 ns, corresponding to a fundamental frequency of 13.6 MHz. In contrast, the repetition period for mode 2 ($OAM_{01}$) is variable but consistently shorter than that of mode 1. Fig. 4a displays the pulse trains and the beam profiles for samples I, II, III and IV, with $\Delta t$ values of 1713 ps, 350 ps, 64 ps, and 0 ps, respectively. The sample I lies within the "Topological" region of the frequency phase diagram under high-frequency mode-locking (Fig. 4d). Fig. 4d illustrates the frequency phase diagram of the laser, which is categorized into three distinct regions: "Topological", "Topological + BEC", and "Trivial". The sample I exhibits a pulse train with uniform intensity distribution, which is consistent with the simulated results (panel 1 in Fig. 3a) for relatively large $\Delta t$. The RF spectrum of sample I (Fig. 4b) shows a free-spectral range (FSR) of 584.2 MHz, adhering to the relation $f = 1/\Delta t$.

Samples II and III fall within the "Topological + BEC" region of the frequency phase diagram. Their pulse trains in Fig. 4a display nested soliton characteristics, where discrete pulses with high harmonic-order frequencies are modulated by a low-frequency envelope at the fundamental repetition rate. This behavior aligns with the simulated results in the groups 2 and 3 of Fig. 3b. The envelope soliton repeats at the fundamental period of 73.5 ns, while the spacing between adjacent pulses equals $\Delta t$ (Supplementary Fig. 15). The nested soliton states are further verified by the RF spectra in Figs. 4b and 4c. Where the high frequency follows $f = 1/\Delta t$, and the low frequency corresponds to the fundamental frequency of approximately 13.6 MHz. The RF spectrum exhibits a finely nested structure (inset of Fig. 4d), which is measured near the first high-frequency peak in panel 2 of Fig. 4b for sample II. This structure features a central high frequency at 2842.4 MHz (marked by a solid black circle), surrounded by satellite lines spaced at approximately 13.6 MHz intervals, matching the fundamental frequency of the single-ring cavity. The formation of nested solions in this topological laser is associated with a BEC-like effect, where increased soliton density leads to condensation and the emergence of an envelope soliton within the cavity. Remarkably, the RF spectra evolve as $\Delta t$ decreases from 350 ps to 0 ps, and the high-harmonic frequencies become more pronounced and eventually form a uniform pattern sequence at $\Delta t$ = 0 ps (Fig. 4c).

Additionally, the fundamental longitudinal mode positions, marked by black rectangles in Fig. 4c, undergo a linear redshift with decreasing $\Delta t$, as indicated by the low-repetition-rate curve in the "Topological + BEC" region of Fig. 4d. Although the two single-mode paths differ in length, only one dominant peak appears in the RF spectrum for each case (Supplementary Fig. 17), indicating that the longitudinal mode repetition rates of the two branches are synchronized. This synchronization may result not only from nonlinear compensation of the intermodal group velocity delay but also from crosstalk or mutual injection between the single-mode paths, which further strengthens intermodal nonlinear interactions in the multimode fiber.

The time-integrated beam profiles (inset of Fig. 4a), captured via a CCD camera, reveal an eccentric disk with an intensity singularity. Mode decomposition at the spectral peak (Supplementary Figs. 15 and 16) shows intensity and phase distributions resembling an eccentric disk with an off-center singularity, and a modal power composition dominated by $OAM_{00}$ and $OAM_{01}$. Similar characteristics are observed in the other samples.

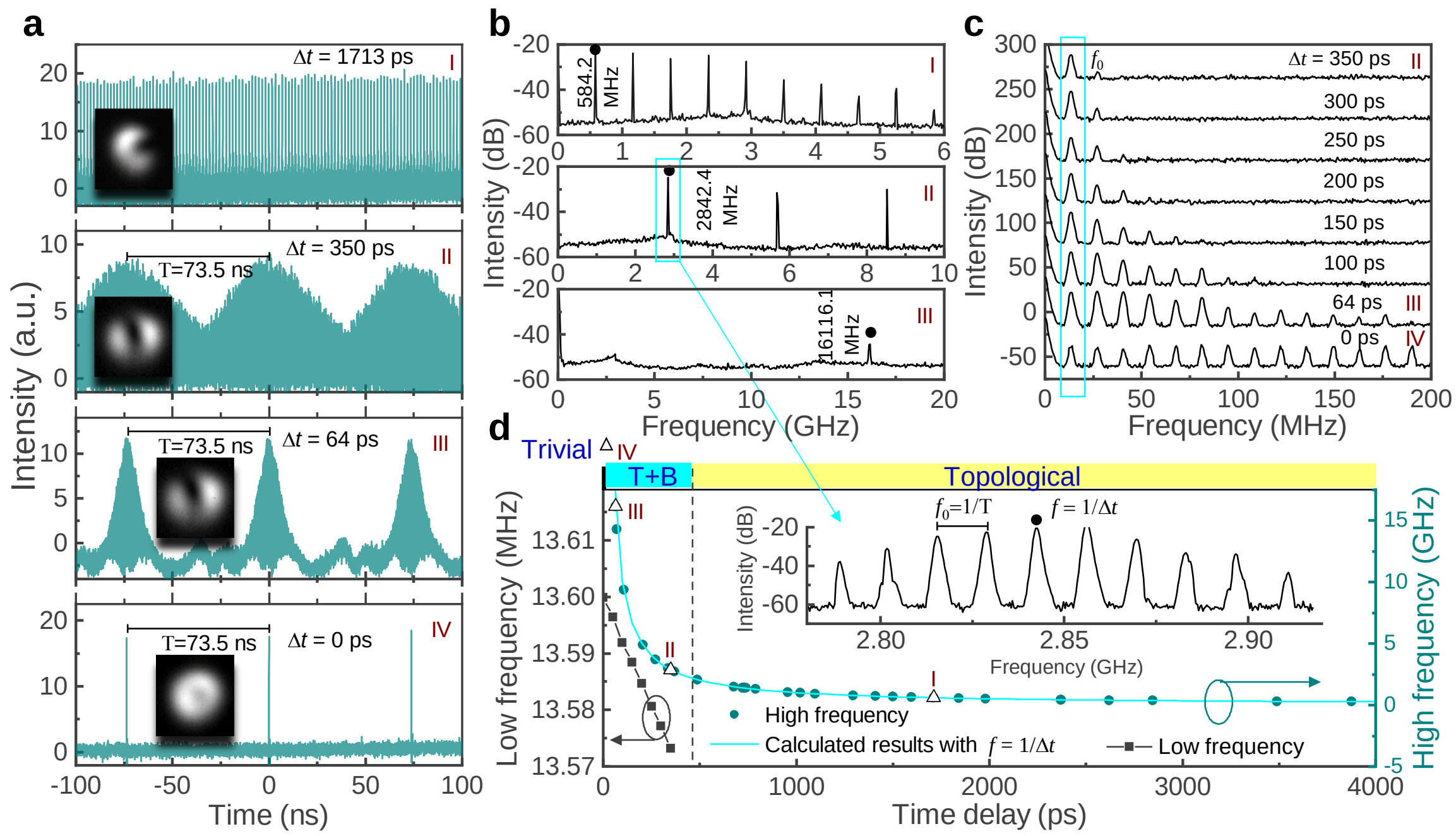

**Fig. 4| Experimental characterization of the frequency phase diagram in the topological spatiotemporal mode-locked fiber laser. a**, Measured pulse trains for four samples I to IV with time delays of 1713 ps, 350 ps, 64 ps, and 0 ps, corresponding to distinct lasing states in the frequency phase diagram shown in **d**. Insets display the respective time-integrated beam profiles. **b**, **c**, RF spectra for samples at specific time delays. **d**, Frequency phase diagram identifying three characteristic regions: "Topological", "T+B" ("Topological + BEC"), and "Trivial". The inset presents a high-resolution spectrum measured near the first high-frequency peak for sample II.

## Discussion and Outlook

The presented laser system, based on an MDM architecture, enables independent control over transverse modes and thereby offers unprecedented flexibility in shaping spatiotemporal wave packets—a capability seldom achieved in conventional spatiotemporal mode-locked fiber lasers. Unlike the laser systems that rely on the collective interplay of dispersion, nonlinearity, and spatial filtering to establish mode-locking, our design incorporates a topologically engineered structure within an MDM framework, providing an effective strategy to realize mode-locking with selectively addressable transverse modes. Topological spatiotemporal solitons in the laser cavity are confined within periodic potential wells and mutually linked through dissipative coupling. The laser working in the "Topological" and "Topological + BEC" states exhibit significant stability against the environmental perturbations, compared to that in "Trivial" states (Supplemental Section 5). Environmental disturbances such as temperature fluctuations and mechanical vibrations may be further suppressed via methods such as the phase-locked loop stabilization, to achieve high-precision spatiotemporal frequency combs [53].

To further improve the controllability and structural precision of spatiotemporal light fields, the number of single-mode branches in the MDM configuration can be expanded.

This allows not only more complex field manipulation but also higher total output power. As a proof of concept, we numerically investigate a topological spatiotemporal mode-locked fiber laser with a six-branch MDM structure under both ideal and disordered conditions (intensity and phase noise), confirming its robust performance (Supplementary Fig. 18). The scalability of the MDM configuration could be further enhanced using cascaded fiber-based directional modal couplers, fiber-based photonic lanterns, or integrated waveguide-based photonic chips. Although expanding the number of single-mode paths enlarges the optical transfer matrix and increases control complexity, this can be alleviated by embedding active components inside the fiber cavity and deploying programmable control algorithms. The present system employs a graded-index multimode fiber as the main cavity medium, which offers relatively low modal dispersion compared to step-index multimode fibers. Future work may explore step-index passive or active multimode fibers with larger modal dispersion, and assess the capability of the MDM architecture to compensate for such dispersion—thus extending the tunability of spatiotemporal mode-locked lasers. Moreover, while the present laser operates in the anomalous dispersion regime, future designs operating in the other dispersion regime could open pathways to study dispersion-managed or all-normal soliton dynamics.

The delay between different single-mode branches of MDM configuration induces periodic potentials along the MMF (longitudinal direction), where the spatiotemporal solitons are assembled. In addition, the graded-index distribution of the MMF induces the parabolic potential in the cross-section (transverse direction). Therefore, the robust of the spatiotemporal soliton formed in the studied laser can be attributed to the 3D potential in both the longitudinal and transverse directions. In this sense, the studied laser system is a practice platform as described in the theoretical works about BEC for robust high-dimensional soliton confinement. The theoretical works indicate that the gauge potentials [54] and 3D parabolic potential [55-57] favor robust high-dimensional high-order solitons. Remarkably, the emergence of nested solitons and the associated temporal envelope narrowing in the studied laser reflect a relation to the BEC in matter-wave systems.

**Acknowledgments:** We thank Dr. Liang Fang from the Hunan University and Dr. Yize Liang from Xidian University for valuable discussions.

**Funding:**

National Natural Science Foundation of China (NSFC) (62125503, 62261160388); Natural Science Foundation of Hubei Province of China (2023AFA028); Hubei Optical Fundamental Research Center (HBO2025TQ004); Chinese Postdoctoral Science Foundation (2022M711242); Funding of the Postdoctoral Innovation Research Post of Hubei Province of China.

**Author contributions:** J.W. and X.L. developed the concept and conceived the experiment. X.L. performed the theoretical analysis and carried out the experiment. X.L. and J.W. analyzed the simulation and experimental data. Y.Z. M.X. B.H. C.C. and J.L. provided technical supports. X.L. wrote the manuscript. J.W. revised and finalized the paper. J.W. supervised the project. All authors discussed the results and contributed to the reviewing of the paper.

**Competing interests:** The authors declare no competing interests.

Supplementary Materials for

# Topological Spatiotemporal Mode-Locked Lasers

Xiaokang Lian[1,2,3], Yuchen Zhang[1,2,3], Mutian Xu[1,2,3], Bing Han[1,2,3], Chengkun Cai[1,2,3], Jun Liu[1,2,3], Jian Wang[1,2,3*]

[1]Wuhan National Laboratory for Optoelectronics and School of Optical and Electronic Information, Huazhong University of Science and Technology, Wuhan 430074, Hubei, China
[2]Hubei Optical Fundamental Research Center, Wuhan 430074, Hubei, China
[3]Optics Valley Laboratory, Wuhan 430074, Hubei, China
* Corresponding Author: J.W. (jwang@hust.edu.cn)

## Contents

# 1 Experimental setup

## 1.1 Details of experimental setup

The proposed laser cavity schematically shown in Fig. 1 comprises a main loop and a mode-division-multiplexing (MDM) sub-circuit. The main loop consists of a graded-index multimode fiber (MMF, OM3) and a home-made multimode fiber coupler (MM OC) whose 10% port serves as the primary output. The MDM section integrates a mode demultiplexer (DEMUX) and a mode multiplexer (MUX), interconnected by two single-mode fiber branches.

The OM3 MMF supports hundreds of transverse modes, but only the lowest two ($OAM_{00}$ and $OAM_{01}$) are selectively accessed through the single-mode branches. Both the DEMUX and MUX are directional couplers fabricated via the fused-biconical-taper technique using a standard single-mode fiber (SMF-28) and an OM3 fiber, as illustrated in Fig. 2. In the DEMUX [Fig. 2(a)], the $OAM_{01}$ is efficiently coupled into the fundamental mode of the SMF-28, whereas the fundamental mode $OAM_{00}$ and other higher-order modes remain guided within the OM3 fiber. The OM3 and SMF ports on the right side of the DEMUX are on-axis spliced to the upper and lower single-mode branches (branch-1 and branch-2), respectively. Owing to on-axis excitation, the $OAM_{00}$ in the OM3 fiber exhibits the highest coupling efficiency with branch-1. The optical fields from branch-1 and branch-2 are subsequently recombined into the $OAM_{00}$ and $OAM_{01}$ modes of the MMF through the MUX, as depicted in Fig. 2(b).

Each single-mode branch contains a polarization-independent isolator (ISO), 980/1550 nm wavelength-division multiplexers (WDMs), erbium-doped fibers (EDFs), 1% output single-mode couplers (SM OC-1 and SM OC-2), three-paddle polarization controllers (PC-1 and PC-2), and saturable absorbers (SA-1 and SA-2). The EDFs are optically pumped by 980 nm single-mode laser diodes (LDs). SAs are implemented by inserting a short section of OM3 fiber between two SMFs, exploiting the nonlinear multimode-interference effect. The ISOs ensure unidirectional operation of the ring cavity. A tunable optical time delay line (OTDL) is incorporated in branch-2 to balance the optical path-length difference between the two branches. The OTDL provides a tuning range of 600 ps, while larger delays can be realized by trimming or extending the SMF sections in either branch.

The length of the adopted MMF (commercial OM3 fiber) is $L_{MMF}$ = 2 m, with second-order dispersion parameters $\beta_2$ = -0.0293 ps$^2$/m for $OAM_{00}$ and $\beta_2$ = -0.0294 ps$^2$/m for $OAM_{01}$. The passive fiber and components in the two single-mode branches are based on conventional SMF-28 fiber, with a total length of $L_{smf}$= 13.3 m for each branch and a second-order dispersion parameter $\beta_2$ = -0.0228 ps$^2$/m. The active gain fiber EDF is SM-ESF-7-125 (from Nufern), with a length of $L_{gain}$= 1.5 m and $\beta_2$ = 0.0485 ps$^2$/m. The total cavity length is approximately 15.3 m. The net dispersion for branch-1 and branch-2 is estimated to be approximately -0.2549 ps$^2$ and -0.2551 ps$^2$, respectively, confirming the operation in the anomalous dispersion regime.

Laser outputs are characterized using an optical spectrum analyzer (YOKOGAWA AQ6370C), a radio-frequency spectrum analyzer (Agilent E4407B), an autocorrelator (Femtochrome FR-103 XL), a digital oscilloscope (Keysight DSA-Z 204A), a CCD camera, and optical power meters. The 3D pulse-field measurement and mode decomposition are performed using delay-scanned off-axis digital holography [1].

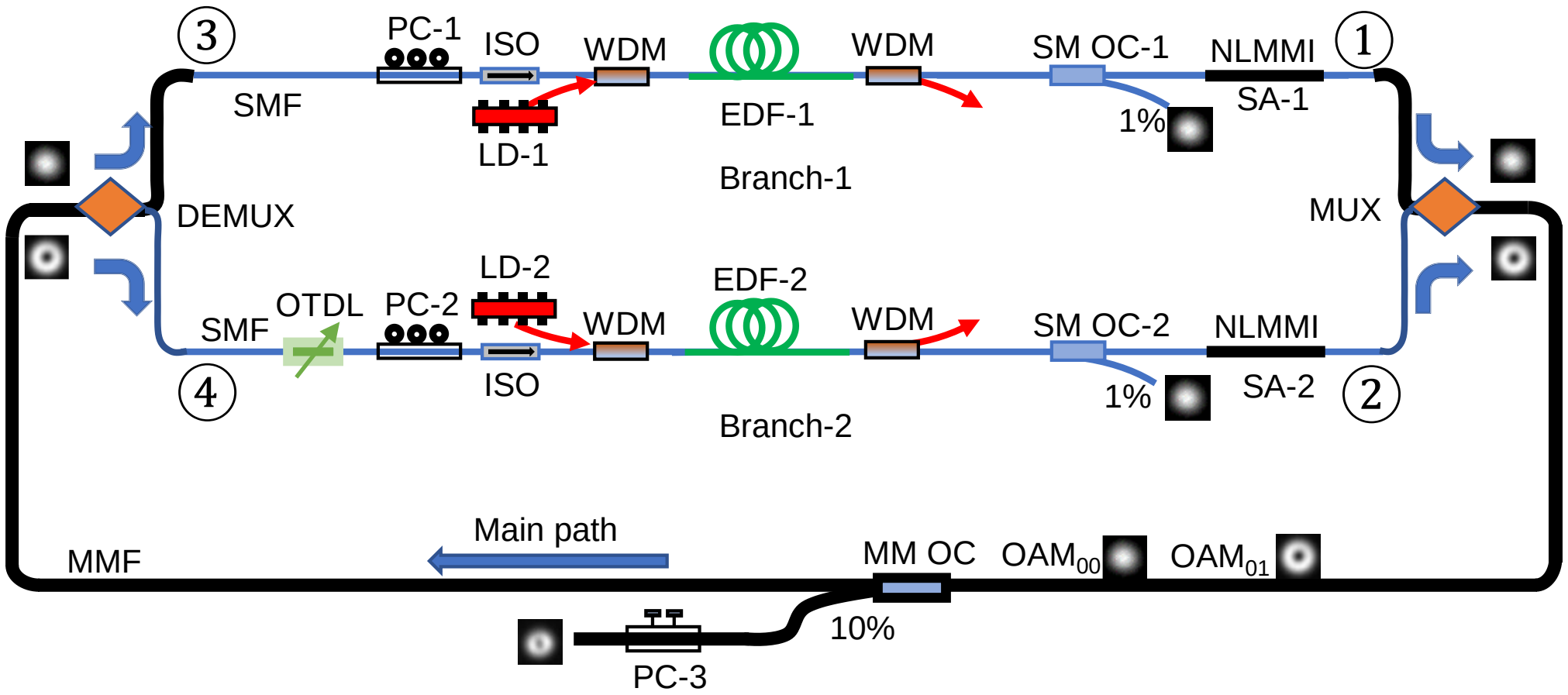

Figure 1: **Configuration of spatiotemporal mode-locked fiber laser based on MDM.** MMF: multimode fiber; SMF: single-mode fiber; MUX: multiplexer; DEMUX: demultiplexer; OTDL: optical time delay line; WDM: wave-division multiplexing; PC: polarization controller; ISO: isolator; LD: 980 nm laser diode; EDF: erbium-doped fiber; SM OC: single-mode optical coupler; NLMMI: nonlinear multimode interference; SA: saturable absorber; MM OC: multimode optical coupler.

### 1.2 The demultiplexer and the multiplexer

The DEMUX and MUX are implemented as directional modal couplers (Fig. 2), fabricated by fusing and tapering a single-mode fiber (SMF-28) with an OM3 multimode fiber under intermodal and vectorial phase-matching conditions [2–4]. The vectorial coupling relationships between the guided modes in the SMF and OM3 are illustrated in the upper panel of Fig. 2(c). Specifically, the x-polarized fundamental mode ($HE_{11}^{x}$) in the SMF couples to the $TE_{01}$ and $HE_{21}^{even}$ modes in the OM3, whereas the y-polarized $HE_{11}^{y}$ couples to the $TM_{01}$ and $HE_{21}^{odd}$ modes. Throughout this work, the orbital angular momentum (OAM) basis is employed in both simulations and experiments due to its distinct azimuthal phase distribution. Variations in the birefringence of the SMF modify the phase and polarization of the fundamental mode, thereby altering the vectorial composition and phase of the coupled modes in the OM3. Different combinations of $TE_{01}$, $HE_{21}^{even}$, $TM_{01}$, and $HE_{21}^{odd}$ form several superpositions corresponding to linearly polarized (LP) or OAM states [5], as summarized in the lower expressions of Fig. 2(c).

The coupling efficiency of the directional modal couplers is sensitive to both the polarization and spatial profiles of the interacting modes [4]. In addition, polarization states are influenced by the birefringence of the fiber paths; variations in birefringence induce wavelength shifts through polarization-dependent spectral filtering, which arises from the phase difference between two orthogonal polarization components of the same transverse mode [6]. Consequently, the insertion loss of the DEMUX and MUX depends on the polarization state, ultimately influencing the overall cavity performance.

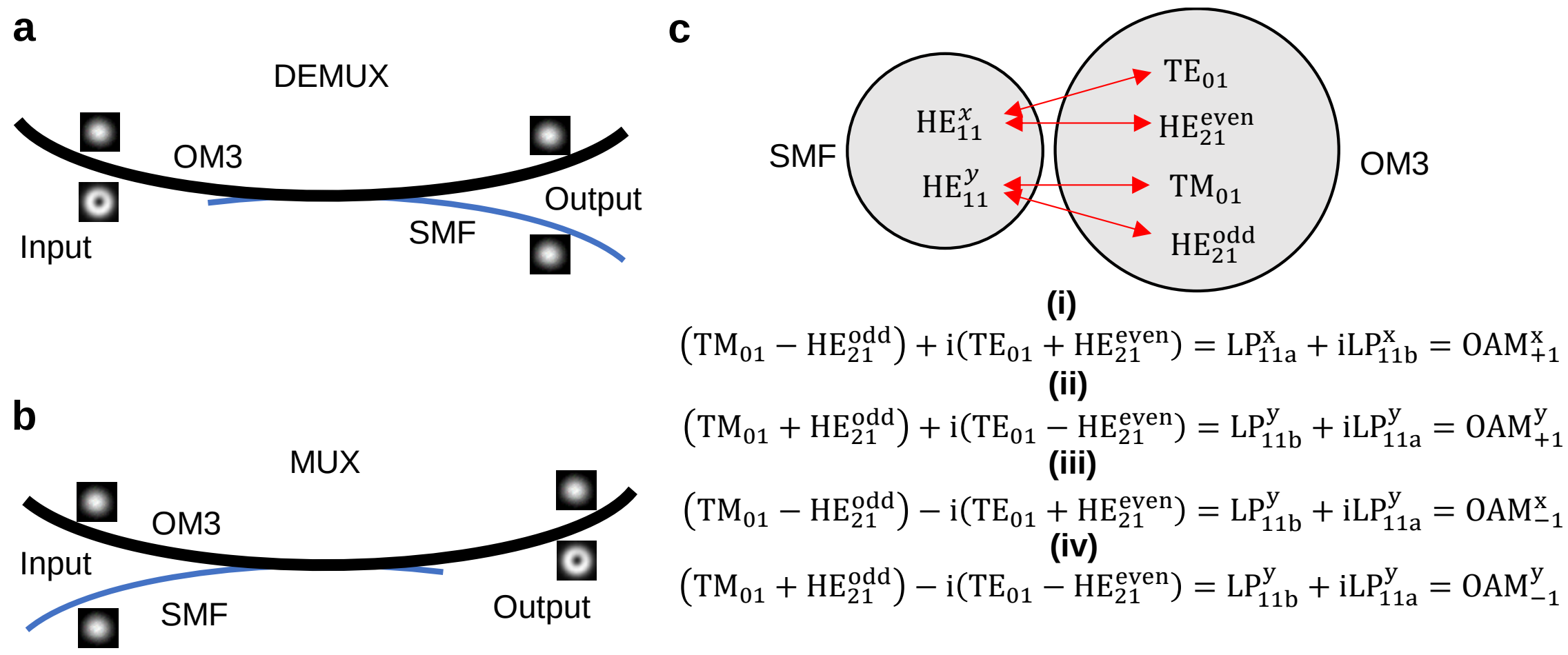

Figure 2: **The configuration of the demultiplexer and the multiplexer.** (**a**) shows the demultiplexer (DEMUX) and (**b**) the multiplexer (MUX). (**c**) Upper: the vectorial coupling relationship between modes in the SMF and the OM3; Lower: the mode conversion relationship in different mode basis. LP: the linearly polarization mode, OAM: the orbital angular mode.

### 1.3 The crosstalk of the laser system

Intermodal crosstalk is an inherent phenomenon in multimode fiber systems, particularly within the MMF, multimode coupler (MM OC), MUX, and DEMUX. Such crosstalk can arise from fabrication imperfections, fusion-splicing errors, fiber bending or twisting, and polarization-dependent loss. Moreover, tuning the polarization controllers (PCs) modifies the birefringence of the single-mode fibers and thereby changes the polarization state of the laser cavity. These variations affect the mode-coupling process in the MUX and DEMUX through the vectorial mode-matching mechanism [3], ultimately inducing polarization-dependent intermodal crosstalk.

Figure 3(a) highlights the section of the laser cavity (circled regions ①–④ in Fig. 1) where crosstalk is most likely to occur, encompassing the MUX, DEMUX, MMF, and MM OC. The optical path from port 1 to port 3 corresponds to single-mode branch 1, supporting the $\mathrm{OAM}_{00}$ mode, whereas the path from port 2 to port 4 corresponds to branch 2, supporting the $\mathrm{OAM}_{01}$ mode. During measurements, a continuous-wave (CW) single-mode laser at 1560 nm was injected into port 1 (or port 2) through a three-paddle polarization controller used to vary the input polarization. The output powers at port 3 and port 4 were then recorded as the PC paddles were rotated. As shown in Fig. 3(b), when light was launched along the $\mathrm{OAM}_{00}$ path (port 1 → port 3), a weak signal was also detected at port 4, indicating polarization-dependent crosstalk. The crosstalk level varied between approximately 10 dB and 17 dB with changes in the polarization state. A similar polarization-dependent behavior was observed when launching light along the $\mathrm{OAM}_{01}$ path (port 2 → port 4), where crosstalk appeared in the port 2–port 3 channel.

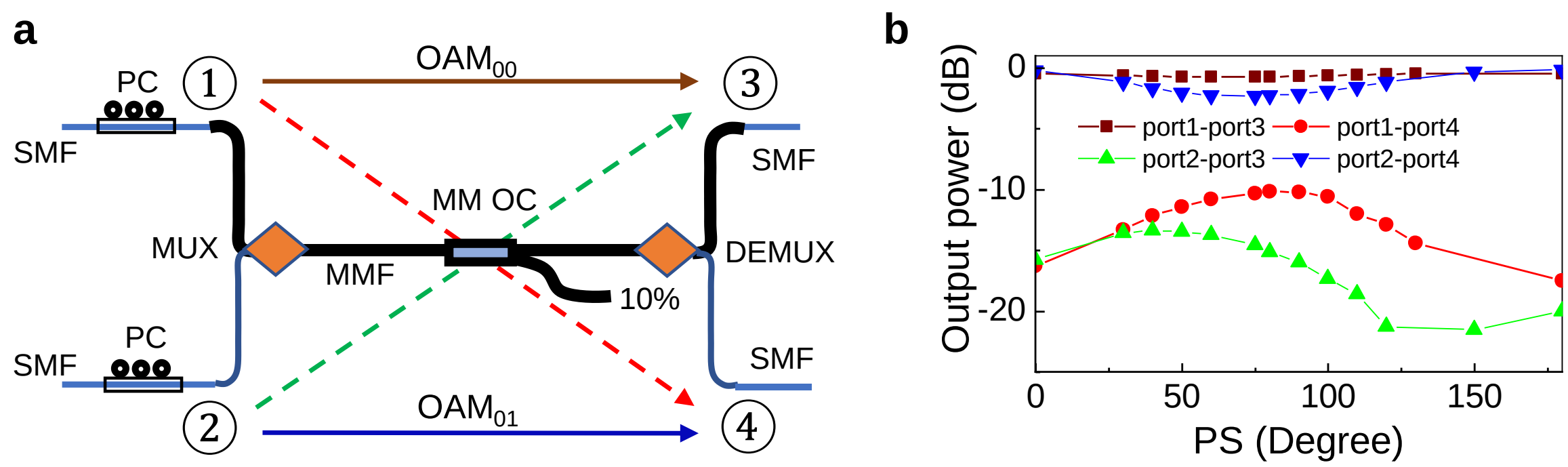

Figure 3: **Measurements of the crosstalk between two single-mode paths.** (**a**) The measurement configuration includes MUX, DEMUX, MMF and MM OC. (**b**) The output power as a function of polarization states (PS) modulated by the polarization controller (PC) in the input port. In (a), the real brown arrow indicates the path of $OAM_{00}$ from port1to port3 while the dashed red arrow indicates the crosstalk of $OAM_{00}$ to $OAM_{01}$ from port1 to port4. The real blue arrow indicates the path of $OAM_{01}$ from port2 to port4 while the dashed green arrow indicates the crosstalk of $OAM_{01}$ to $OAM_{00}$ from port2 to port3.

## 2 Photonic lattices and simulation models

### 2.1 Photonic lattices in topological passively mode-locked fiber lasers

As discussed in the main text, the mode-locked fiber lasers based on the two-branch MDM are topologically equivalent to a coupled laser oscillator model. As illustrated in Main Text Fig. 1 and Fig. 4(a), the oscillation periods of the upper and lower oscillators are denoted by $T_1$ and $T_2$, corresponding to fundamental frequencies $f_1$ and $f_2$, respectively. The lasing period and frequency of the spatiotemporal solitons are $T$ and $f$, which are related by:

$$\begin{cases} f_1 = \frac{1}{T_1} \\ f_2 = \frac{1}{T_2} \\ f = \frac{1}{T} = \frac{1}{T_1 - T_2} \end{cases} \tag{1}$$

For a stable harmonic mode-locked laser, the two oscillators must be frequency-synchronized. Their harmonic orders differ due to the distinct lengths (or fundamental periods) of the individual oscillators. Denoting the harmonic orders of the upper and lower oscillators as $N_1$ and $N_2$, the relationship between these quantities is expressed as:

$$\begin{cases} f = N_1 f_1 \\ f = N_2 f_2 \end{cases} \tag{2}$$

Then, it can be deduced from Eq. (1) and Eq. (2) that $N_1$-$N_2$=1. In other words, there is one more soliton in the long upper oscillator than that in the lower oscillator.

The soliton dynamics within the laser cavity are analyzed using a two-dimensional lattice representation, as shown in Figs. 4(b) and 4(c). Pulses in the upper and lower oscillators are denoted by $u_n^m$ and $v_n^m$, respectively, where $m$ indexes the long-term time evolution and $n$ the short-term time step (or lattice position along the oscillator). At the coupling site (highlighted in orange, representing the combined MMF, MUX, and DEMUX), the pulse comprises contributions from both transverse modes, $u_n^m$ and $v_n^m$. From a given coupling site, $u_n^m$ travels along the upper long oscillator while $v_n^m$ propagates along the lower short oscillator, reaching distinct lattice positions after one long-term period due to the inter-path time delay. The interaction between the two modes at the coupling site constitutes dissipative coupling, arising from two mechanisms. First, the time delay introduces a reservoir effect, whereby a portion of the pulse energy remains in the delay line and coupling occurs between the residual portions. Second, the directional couplers and transverse-mode interactions in the MMF mediate coupling via a third mode (Section 1.2), which inevitably induces radiation leakage.

If set the upper oscillator as the main cavity and the lower oscillator is considered as a delay line reservoir, the pulse lattice in two dimensions with $n\Delta t$ as the lattice position and $mT_1$ as the long-term time step, shown in Fig. 4(b). In this case, the mathematical foundation of the pulse lattice is given as:

$$\begin{cases} u_n^{m+1} = \sqrt{1-\rho}u_n^m + \sqrt{\rho}v_{n+1}^m \\ v_n^{m+1} = \sqrt{\rho}u_n^m + \sqrt{1-\rho}v_{n+1}^m \end{cases} \tag{3}$$

It is indicated from the first equations that the pulse $u_n^{m+1}$ is partial from $\sqrt{1-\rho}u_n^m$ and partial from the $\sqrt{\rho}v_{n+1}^m$, implying that the distribution of modal pulses $u_n$ are fixed in the periodic round trips, and from the second equations that the modal pulses $v_n$ is moving as a unidirectional flow. The distribution of $u_n$ can be described as the unidirectional Hatano-Nelson lattice as shown in Main Text Fig. 1(e), which is characterized by the periodic pulses with nearest-neighbor nonreciprocal dissipative couplings.

Similarly, if set the lower oscillator as the main cavity and the upper as the reservoir, the mathematical foundation of the pulse lattice is given as follows,

$$\begin{cases} u_n^{m+1} = \sqrt{1-\rho}u_{n-1}^m + \sqrt{\rho}v_n^m \\ v_n^{m+1} = \sqrt{\rho}u_{n-1}^m + \sqrt{1-\rho}v_n^m \end{cases} \tag{4}$$

It can be deduced that the distribution of $v_n$ can also be described as anther unidirectional Hatano-Nelson lattice, as indicated by Fig. 4(c).

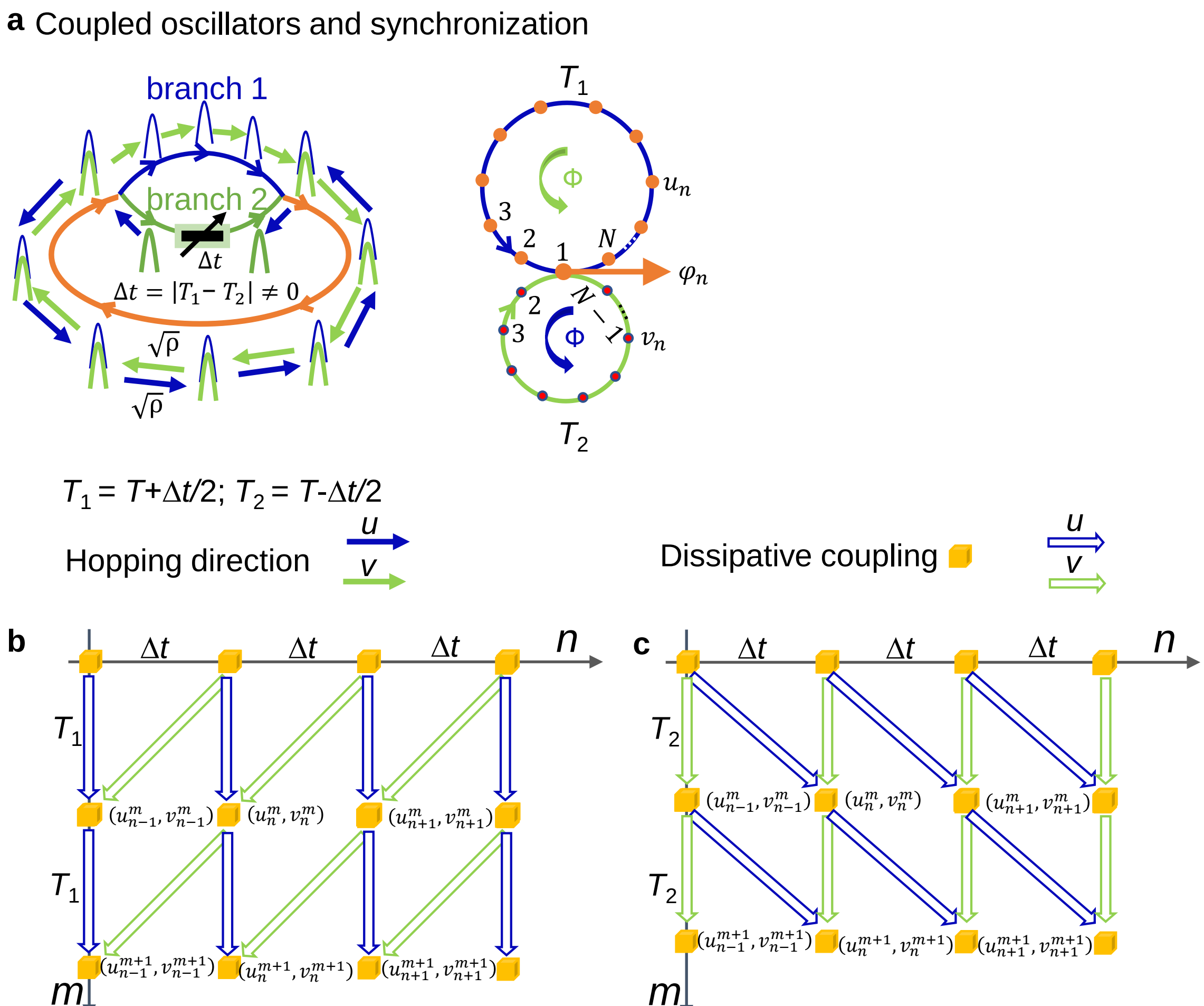

$$\begin{cases} u_n^{m+1} = \sqrt{1-\rho}\,u_n^m + \sqrt{\rho}\,v_{n+1}^m \\ v_n^{m+1} = \sqrt{\rho}\,u_n^m + \sqrt{1-\rho}\,v_{n+1}^m \end{cases}$$

$$\begin{cases} u_n^{m+1} = \sqrt{1-\rho}\,u_{n-1}^m + \sqrt{\rho}\,v_n^m \\ v_n^{m+1} = \sqrt{\rho}\,u_{n-1}^m + \sqrt{1-\rho}\,v_n^m \end{cases}$$

Figure 4: **Pulse lattice in the topological mode-locked laser cavity with $\Delta t \neq 0$.** (**a**) Distribution of pulses in the laser cavity, left is shown in MDM configuration while right is shown in coupled oscillator configuration. (**b**) Pulse lattice in two dimensions with $n\Delta t$ as the lattice position and $mT_1$ as the time step. (**c**) Pulse lattice in two dimensions with $n\Delta t$ as the lattice position and $mT_2$ as the time step.

### 2.2 The mean-field master function

The topological mode-locked fiber laser based on MDM can be simplified as coupled oscillators with different delays. The coupled master function of mode-locking in multiple coupled oscillators can be derived from the passive fast saturable absorber model with a delay-coupled term [8,9]:

$$i\frac{\partial \psi^p(T,t)}{\partial (T)} = \epsilon_1^p \frac{\partial^2}{\partial t^2}\psi^p(T,t) - \epsilon_2^p|\psi^p(T,t)|^2\psi^p(T,t) + i\epsilon_3^p\psi^p(T,t) + \sum_r^N k_r e^{i\theta_r}\psi^r\left(T - \frac{rT_p}{N}, t\right) \quad (5)$$

where $\psi^p(T,t)$ is for the branch-p, with the period $T_p$, $\epsilon_1^p = i\frac{g(T)}{\Omega_f^2} - D$, $\epsilon_2^p = -i\gamma$, $\epsilon_3^p =$ $g(T) - l - s(t)$. $\psi^r$ is for the branch-r with a delay of $\frac{rT_p}{N}$ to the branch-p. '$r$' is in integer. '$t$' denotes the transverse position in the topological lattice while '$T$' the longitude time. The last term $\sum_r^N k_r e^{i\theta_r}\psi^r\,(T - \frac{rT_p}{N}, t)$ in the Eq. (5) is corresponding to multiple reservoir branches, where $k_r$ denotes the coupling efficient, $\theta_r$ is the accumulated phase difference, '$T_p$' denotes the fundamental period, and '$N$' the harmonic order.

Since the effect of delay is to induce nonreciprocal dissipative coupling, and the coupled oscillators can be understood based on the unidirectional Hatano-Nelson lattice model shown in main text. In order to investigate the dynamics of the laser based on the two-branch MDM, the governing equation can be formulated as following:

$$i\frac{\partial\psi(T,t)}{\partial T} = \epsilon_1\frac{\partial^2}{\partial t^2}\psi(T,t) - \epsilon_2|\psi(T,t)|^2\psi(T,t) + i\epsilon_3\psi(T,t) + k\psi(T,t-\Delta t) \quad (6)$$

with the last term of $k\psi(T,t-\Delta t)$, and $\Delta t = \frac{T_{RT}}{N}$. The Eq. (6) without the last term is a complex Ginzburg-Landau equation (CGLE) with dispersion ($\mathcal{D}$, the first term on the right hand), the nonlinear saturable absorber and materials ($\mathcal{N}$, the second term) and gain and loss ($\mathcal{G}$, the third term) [8]. The last term is the intracavity coupling term, resulted from the delay-line reservoir [10].

The evolution of gain and the saturable absorber in this model are respectively:

$$g(T) = \frac{g_{ss}}{1+\int dt|\psi(T,t)|^2/E_{sat}} \quad (7)$$

where $g_{ss}$ is the small signal gain. $E_{sat}$ is the saturated energy, which represents the pump power. $s(t)$ is the transfer function of the saturable absorber expressed as:

$$s(t) = \frac{q_0}{1+|\psi(T,t)|^2/P_{sat}} \quad (8)$$

where $q_0$ is the modulation depth and $P_{sat}$ is the saturation power.

### 2.3 The tight-binding model

The delay-induced periodic lattice potential (Section 3), effectively decomposes the mean-field master equation into a set of weakly coupled nodes. Assuming $N$ pulses for $N^{\text{th}}$-order harmonic mode-locking are tightly bound at equidistant positions, the cavity can be divided in to $N$ sections, each described by a pulse $\psi_n(T,t)$. The evolution of the $n^{\text{th}}$ pulse follows:

$$\frac{\partial\psi_n(T,t)}{\partial T} = \epsilon_1\frac{\partial^2}{\partial t^2}\psi_n(T,t) - \epsilon_2|\psi_n(T,t)|^2\psi_n(T,t) + i\epsilon_3\psi_n(T,t) \quad (9)$$

This equation has a steady-state soliton solution [9]. Projecting the pulse dynamics onto their central manifolds (peak amplitudes) and incorporating the nearest-neighbor intracavity coupling would yield a tight-binding model, described by a discrete complex Ginzburg-landau equation (dCGLE):

$$\frac{\partial\varphi_n}{\partial T} = \epsilon_1\frac{\partial^2}{\partial t^2}\varphi_n(T,t) - \epsilon_2|\varphi_n(T,t)|^2\varphi_n(T,t) + i\epsilon_3\varphi_n(T,t) + k_L\varphi_{n+1} + k_R\varphi_{n-1} \quad (10)$$

The last term is omitted for the unidirectional Hatano-Nelson lattice model [Main Text Fig. 1(e)] is obtained:

$$\frac{\partial\varphi_n}{\partial T} = \epsilon_1\frac{\partial^2}{\partial t^2}\varphi_n(T,t) - \epsilon_2|\varphi_n(T,t)|^2\varphi_n(T,t) + i\epsilon_3\varphi_n(T,t) + k\varphi_{n+1} \quad (11)$$

The evolution of gain and the saturable absorber in this model are respectively:

$$g(T) = \frac{g_{ss}}{1+\sum_n|\varphi_n|^2 * 2 * \tau_n/E_{sat}}$$

$$s(t) = \frac{q_0}{1 + |\varphi_n|^2/P_{sat}}$$

**2.4 Lumped element model**

The MDM-based multimode fiber laser can be simulated using a lumped propagation model, where each cavity component is represented by its transmission functions. Light propagation in the SMF, EDF and MMF is simulated using the coupled generalized multimode nonlinear Schrödinger equation (GMMNLSE) [11]:

$$\frac{\partial A_p(z,t)}{\partial z} = i\left(\beta_0^{(p)} - \beta_0^{(1)}\right)A_p - \left(\beta_1^{(p)} - \beta_1^{(1)}\right)\frac{\partial A_p(z,t)}{\partial z} - i\frac{\beta_2^{(p)}}{2}\frac{\partial^2 A_p(z,t)}{\partial t^2} + \frac{g}{2}A_p(z,t) + i\frac{n_2\omega_0}{c}\sum_{l,m,n} S_{plmn}^K A_l A_m A_n^* \quad (12)$$

where $A_p(z,t)$ is the electric field envelope for the mode $p$, $\beta_n^{(p)}$ is the $n^{th}$-order dispersion for mode $p$, and $S_{plmn}^K$ is the nonlinear coupling coefficient for the Kerr effect. $g$ is the gain coefficient of EDF, and $g = 0$ for SMFs and MMFs. The $g$ term is computed in the frequency domain as:

$$g = \frac{g_{ss}}{1 + \int dt\left|A_p(z,t)\right|^2/E_{sat}} f(\omega)$$

where $g_{ss}$ is the small signal gain, and the gain spectral bandwidth is denoted by the Lorentz function $f(\omega)$ as follows:

$$f(\omega) = \frac{\Delta\omega}{[(\omega - \omega_c)^2 + (\Delta\omega/2)^2]}$$

where, $\Delta\omega$ is the bandwidth of the Lorentzian gain profile and $\omega_c$ is the central frequency.

The saturable absorber based on nonlinear multimode interference effect is modeled by a general transfer function as follows:

$$T(t,z) = \sqrt{1 - \frac{q_0}{1 + |A(t,z)|^2/P_{sat}}}$$

where $q_0$ is the modulation depth and $P_{sat}$ is the saturation power.

The intermodal crosstalk, arising from fabrication imperfections, fusion splicing errors, fiber bending, twisting, and polarization-dependent loss, is present in the multimode fiber system, including MMF, MM OC, MUX and DEMUX, as discussed in “Materials and Methods”. This crosstalk facilitates mutual injection between the two single-transverse-mode paths, synchronizing the transverse-mode solitons. In simulations, the total crosstalk effect from all cavity components is modeled by a 2×2 matrix as follows:

$$M = \begin{bmatrix} \sqrt{1-\rho} & \sqrt{\rho} \\ \sqrt{\rho} & \sqrt{1-\rho} \end{bmatrix}$$

where the crosstalk coupling coefficient is set to $\rho = 0.1$ to accelerate the convergence.

The optical fiber time delay line is modeled by applying a corresponding phase delay in the frequency domain:

$$A(t,z) \rightarrow \text{ifft}\{\text{fft}\,[A(t,z)]\exp\,(-i\omega\Delta t)\}$$

where $\Delta t$ is the time delay between the two single-mode fiber branches.

In simulations, light propagates sequentially through the passive SMF, active EDF, SA, MUX, MMF, MM OC, DEMUX, and back to the passive SMF. The fiber lengths and dispersion parameters match the experimental values. For the EDF, the gain bandwidth is $\Delta\omega$ = 5.7 THz and the small-signal gain is $g_{ss}$ = 3 (1/m). The total crosstalk coefficient $CT$ in the laser cavity is set as 0.1 to accelerate the convergence. For the simulation of "Trivial" spatiotemporal mode-locking, the time delay is of $\Delta t$ = 0 ps. The saturable energy $E_{sat}$ of the EDF is variable (the larger $E_{sat}$, the larger pump power). The modulation depth and saturable power of the SA are $q_0$= 0.3 and $P_{sat}$ = 100 W, respectively. For the simulation of "Topological" and "Topological + BEC" spatiotemporal mode-locking, the parameters are $q_0$= 0.3, $P_{sat}$ = 20 W, and $\Delta t$ =$\Delta T$+GVD $\times$ $L_{MMF}$, where GVD is the group velocity difference and $\Delta T$ is the soliton period. The value of $E_{sat}$ varies case by case as indicated in the main test.

## 3 Delay induced periodic potential in the topological soliton lattice

In this section, the collective coordinate method is used to deduce the delay induced periodic potential in the topological lattice formed in our laser system. In Eq. (6), the field $\psi(T, t-\Delta t)$ at the delay time of '$t-\Delta t$' is equivalent to the delay field at the normal time of '$t$' with a form of $e^{i\theta_0}\psi(T-\Delta T, t)$. $\theta_0$ is the accumulated phase due to the time delay. Therefore, the equation can be converted into

$$i\frac{\partial\psi(T,t)}{\partial T}-\epsilon_1\frac{\partial^2}{\partial t^2}\psi(T,t)-\epsilon_2|\psi(T,t)|^2\psi(T,t)=i\epsilon_3\psi(T,t)+\ ke^{i\theta_0}\psi(T-\Delta t,t) \tag{13}$$

which is a kind of delayed GLE [9,12], supporting periodic soliton solution.

The three terms on the left hand of Eq. (13) denote the time derivative, the dispersion and the nonlinear terms, respectively, which is a main Schrödinger function for the transited soliton (usually generated in a fiber with anomalous dispersion).

$$i\frac{\partial\psi(T,t)}{\partial T}-\epsilon_1\frac{\partial^2}{\partial t^2}\psi(T,t)-\epsilon_2|\psi(T,t)|^2\psi(T,t)=0 \tag{14}$$

The two terms on the right hand of Eq. (13) denote the gain and loss, and the coupling terms, which can be considered as the perturbation to the soliton (weak nonlinearity, weak dispersion, good balance between loss and gain) [13-15].

$$R=i\epsilon_3\psi(T,t)+ike^{i\theta_0}\psi(T-\Delta t,t) \tag{15}$$

In order to drive the Lagrangian density $\mathcal{L}$, given by

$$\mathcal{L}=\frac{i}{2}(\psi\psi_T^*-\psi^*\psi_T)-\epsilon_1\left|\frac{\partial\psi}{\partial t}\right|^2+\frac{\epsilon_2}{2}|\psi|^4 \tag{16}$$

the time and space derivatives of $\psi(T,t)$ are computed as following. We adopt the moving soliton ansatz to derive the kinetic term. The wave function is assumed by the Galilean-boosted ansatz for the moving soliton [16]:

$$\psi(T,t)=\ A(\xi)e^{i\Phi(T)+i\frac{v(T)}{2}\xi} \tag{17}$$

where $\xi=t-\tau(T)$, with $A(\xi)$ being the shape of the soliton, $\Phi(T)$ being the global phase, $v(T)$ being the phase gradient (which represents the soliton's instantaneous velocity), and $\tau(T)$ being the soliton's center in collective coordinate method (which we aim to derive the dynamics for).

Form $\xi=t-\tau(T)$, we get $\frac{\partial\xi}{\partial T}=-\dot{\tau}$. And the phase can be denoted as $\theta(T,t)=\Phi(T)+\frac{v(T)}{2}\xi$, and then $\dot{\theta}=\dot{\Phi}+\frac{1}{2}\dot{v}\xi-\frac{v}{2}\dot{\tau}$. The time derivative of $\psi(T,t)$ are as following:

$$\psi_T=\frac{\partial\psi}{\partial T}=\ -\dot{X}A'(\xi)e^{i\theta}+i\,A(\xi)\left(\dot{\Phi}+\frac{1}{2}\dot{v}\xi-\frac{v}{2}\dot{\tau}\right)e^{i\theta}$$

The space derivative of $\psi(T,t)$ are as following:

$$\psi_t = \frac{\partial\psi}{\partial t} = A' e^{i\theta} + i\,\frac{v}{2} A e^{i\theta}$$

and the corresponding expression for $|\frac{\partial\psi}{\partial t}|^2$:

$$\left|\frac{\partial\psi}{\partial t}\right|^2 = A'^2 + \left(\frac{v^2}{4}\right) A^2$$

The time term in the Lagrangian density is:

$$\frac{i}{2}(\psi\psi_T^* - \psi^*\psi_T) = A^2\left(\dot{\Phi} + \frac{1}{2}\dot{v}\xi - \frac{v}{2}\dot{X}\right)$$

We integrate over all space to compute the effective Lagrangian.

$$\begin{aligned} L_{\text{eff}} &= \int_{\frac{T_{RT}}{2}}^{\frac{T_{RT}}{2}} \mathcal{L}dt \\ &= \int_{\frac{T_{RT}}{2}}^{\frac{T_{RT}}{2}} \left[\frac{i}{2}(\psi\psi_T^* - \psi^*\psi_T) - \epsilon_1\left|\frac{\partial\psi}{\partial t}\right|^2 + \frac{\epsilon_2}{2}|\psi|^4\right] d\xi \\ &= \int A^2\left(\dot{\Phi} + \frac{1}{2}\dot{v}\xi - \frac{v}{2}\dot{\tau}\right) d\xi - \epsilon_1 \int \left[A'^2 + \left(\frac{v^2}{4}\right)A^2\right] d\xi + \frac{\epsilon_2}{2}\int A^4 d\xi \\ &= I_0\dot{\Phi} - \frac{v}{2}\dot{\tau} I_0 + \frac{1}{2}\dot{v}Q - \epsilon_1\frac{v^2}{4}I_0 - \epsilon_1 I_1 + \frac{\epsilon_2}{2}I_2 \end{aligned}$$

where $I_0 = \int A^2 d\xi$, $I_1 = \int A'^2 d\xi$, $I_2 = \int A^4 d\xi$, and $Q = \int \xi A^2 d\xi$ . For soliton profiles that are symmetric about $\xi = 0$ (e.g., sech profile), $Q = 0$, and we obtain:

$$L_{\text{eff}} = I_0\dot{\Phi} - \frac{v}{2}I_0\dot{\tau} - \epsilon_1\frac{v^2}{4}I_0 - E_0 \tag{18}$$

where $E_0$ is a constant energy term related to the nonlinear interaction and soliton shape, as shown in Eq. (17).

In the next step, we treat $v(t)$ as an independent generalized coordinate and apply the variational method:

$$\frac{\partial L_{\text{eff}}}{\partial v} = 0$$

which yields the relation:

$$v = -\frac{\dot{\tau}}{\epsilon_1}$$

Substituting this back into Eq. 18, we get:

$$L_{\text{eff}} = I_0\dot{\Phi} + \frac{I_0}{4\epsilon_1}\dot{\tau}^2 - E_0 \tag{19}$$

The standard kinetic energy term is

$$T = \frac{I_0}{4\epsilon_1}\dot{\tau}^2 = \frac{1}{2}M_{\text{eff}}\dot{\tau}^2$$

where $M_{\text{eff}} = \frac{I_0}{2\epsilon_1}$ is the effective mass of the soliton, determined by the soliton profile.

Based on the Euler–Lagrange equation:

$$\frac{d}{dT}\left(\frac{\partial L_{\text{eff}}}{\partial\dot{\tau}}\right) - \frac{\partial L_{\text{eff}}}{\partial\tau} = 0$$

Then, the equation of motion of the center of mass containing inertial terms

$$M_{\text{eff}}\ddot{\tau} = -\frac{\partial U}{\partial\tau} \tag{20}$$

where the right is the derivative of an expected potential due to the perturbation [Eq. (18)], related to the generalized force $F_\tau$. Take the perturbation into the Euler-Lagrange equation [15,16]:

$$\frac{d}{dT}\left(\frac{\partial L_{\text{eff}}}{\partial \dot{\tau}}\right) - \frac{\partial L_{\text{eff}}}{\partial \tau} = 2\text{Re} \int R \frac{\partial \psi^*(T,t)}{\partial \tau} dt \tag{21}$$

The generalized force $F_\tau$ is given by

$$\begin{aligned} F_\tau &= 2\text{Re} \left\{ \int R \frac{\partial \psi^*(T,t)}{\partial \tau} dt \right\} \\ &= 2\text{Re} \left\{ \int \left[ i\epsilon_3 \psi(T,t) + ike^{i\theta_0} \psi(T-\Delta t, t) \right] \frac{\partial \psi^*(T,t)}{\partial \tau} dt \right\} \\ &= 2\text{Re} \left\{ \int i\epsilon_3 \psi(T,t) \frac{\partial \psi^*(T,t)}{\partial \tau} dt + \int ike^{i\theta_0} \psi(T-\Delta t, t) \frac{\partial \psi^*(T,t)}{\partial \tau} dt \right\} \\ &= 2\text{Re} \left\{ \int ike^{i\theta_0} \psi(T-\Delta t, t) \frac{\partial \psi^*(T,t)}{\partial \tau} dt \right\} \end{aligned}$$

where the integration

$$\begin{aligned} &\int \frac{\partial \psi^*(T,t)}{\partial \tau} \cdot \psi(T-\Delta t, t) dt \\ &= -\int_{-\frac{T_{RT}}{2}}^{\frac{T_{RT}}{2}} A'\big(t-\tau(T)\big) e^{-i\theta(T)} \cdot A\big(t-\tau(T-\Delta t)\big) e^{i\theta(T-\Delta t)} dt \\ &= -e^{i[\theta(T-\Delta t)-\theta(T)]} \int_{-\frac{T_{RT}}{2}}^{\frac{T_{RT}}{2}} A'\big(t-\tau(T)\big) \cdot A\big(t-\tau(T-\Delta t)\big) dt \\ &= -e^{i[\theta(T-\Delta t)-\theta(T)]} \int_{-\frac{T_{RT}}{2}}^{\frac{T_{RT}}{2}} A'(\xi) \cdot A\big(\xi-\tau(T-\Delta t)+\tau(T)\big) d\xi \\ &= -e^{i[\Delta\theta]} \int_{-\frac{T_{RT}}{2}}^{\frac{T_{RT}}{2}} A'(\xi) \cdot A(\xi-\Delta\tau) d\xi \\ &= -e^{i[\Delta\Phi]} \int_{-\frac{T_{RT}}{2}}^{\frac{T_{RT}}{2}} A'(\xi+\Delta\tau) \cdot A(\xi) d\xi \\ &= -e^{i[\Delta\theta]} J(\xi) \end{aligned}$$

Here,

$$\Delta\tau = \tau(T-\Delta t) - \tau(T)$$

$$\Delta\theta = \theta(T-\Delta t) - \theta(T)$$

$$J(\xi) = \int_{-\frac{T_{RT}}{2}}^{\frac{T_{RT}}{2}} A'(\xi+\Delta\tau) \cdot A(\xi) d\xi$$

This equation is related to the overlap of the two solitons in spatial region, therefore can be transferred into an equation about their position of $\tau$.

$$\begin{aligned} J(\tau) &= \int_{-\frac{T_{RT}}{2}}^{\frac{T_{RT}}{2}} \frac{\partial}{\partial \tau} A(\xi+\Delta\tau) \cdot A(\xi) d\tau \\ &= \frac{\partial}{\partial \tau} \left[ \int_{-\frac{T_{RT}}{2}}^{\frac{T_{RT}}{2}} A(\xi+\tau) \cdot A(\xi) d\tau \right] \end{aligned} \tag{22}$$

Then:

$$F_\tau = -2ke^{i(\Delta\Phi+\theta_0)}J(\tau)$$

The phase term would not affect the changes of the position, therefore, which could be omitted and keep only the real part. The resulting force is proportional to the overlap of the soliton profiles and has the form of a cosine potential:

$$F_\tau = -2kJ(\tau)\cos(\Delta\Phi + \theta_0) \tag{23}$$

The equation indicates that the displacement of the solitons would induce the force. Since the displacement is small, then the $J(\tau)$ can be assumed as $J_0$, corresponding to the "Topological" mode-locking states, then

$$F_\tau = -2kJ_0\cos(\Delta\Phi + \theta_0) \tag{24}$$

The harmonic field is a periodic type. Therefore, the phase difference is periodic about the spatial coordinate $t$

$$\Delta\Phi = \Phi(T - \Delta t) - \Phi(T) = \omega\tau$$

where $\omega = \frac{2\pi}{\Delta t}$.

$$F_\tau = -2kJ_0\cos(\omega\tau + \theta_0) \tag{25}$$

The periodic potential

$$\begin{aligned} U(\tau) &= -\int_{-\frac{T_{RT}}{2}}^{\frac{T_{RT}}{2}} F_\tau d\tau \\ &= -\int_{-\frac{T_{RT}}{2}}^{\frac{T_{RT}}{2}} kJ_0\cos(\omega\tau + \theta)d\tau \\ &= -\varepsilon\frac{J_0}{k}\sin(\omega\tau + \theta_0) + U_0 \end{aligned} \tag{26}$$

Finally, we achieved the equation of motion for the soliton's center $\tau(T)$, as shown as Eq. (20).

# 4 Additional measurements

## 4.1 The spatiotemporal mode-locking in "Trivial" state

Figure 5 summarizes the spatiotemporal mode-locking characteristics of the laser system with a time delay of $\Delta t$ = 0. Panels 1 in Figs. 5(a) to 5(c) show the optical spectrum, temporal pulse train, and radio-frequency (RF) spectrum for single-mode branch 1 ($OAM_{00}$) at a pump power of p1 = 51 mW, measured at the MM OC. The optical spectrum [Fig. 5(a), panel 1] exhibits a sech-shaped profile with Kelly sidebands, corresponding to characteristics of soliton mode-locking. The full width at half maximum (FWHM) is 4.4 nm from a sech fit. The inset shows the Gaussian beam profile of the output, corresponding to the $OAM_{00}$ transverse mode of the OM3 fiber. The pulse width, determined from the autocorrelation (AC) and sech fitting [Fig. 5(b), inset, panel 1], has an FWHM of 574 fs. The resulted time–bandwidth product (TBP) is 0.31, close to the transform-limited value of 0.315. The RF spectrum [Fig. 5(c), panel 1] displays uniformly spaced longitudinal peaks with a signal-to-noise ratio (SNR) of 47 dB, confirming high-quality soliton mode-locking. For branch 2 ($OAM_{01}$) at p2 = 53 mW, panels 2 in Figs. 5(a) to 5(c) show comparable results: spectral FWHM of 4.5 nm, pulse width of 557 fs, TBP of 0.31, and SNR of 48 dB, indicating high-quality soliton mode-locking too.

When both branches are pumped simultaneously, spatiotemporal mode-locking emerges with $OAM_{00}$ and $OAM_{01}$ modes. Panel 3 in Fig. 5(a) compares the MM OC output spectra with those measured from SM OC-1 and SM OC-2. The spectrum from SM OC-1 (red) shows distortion and a co-lasing CW peak at 1563 nm, distinct from branch-1 alone. SM OC-2 (olive) exhibits periodic modulation, differing from branch-2 alone. The MM OC spectrum (dark cyan) displays multiple broad peaks atop a coarse envelope, indicating a new spectrum formed through self-organization rather than a simple superposition of the individual branch spectra. The inset shows the beam profile, characterized by coherent interference of the two transverse modes.

Spatiotemporal mode-locking formation can be interpreted as synchronization of two transverse-mode solitons in the single-mode branches, evidenced by the overlap of their pulse trains and RF sequences with the multimode main path [panels 3 of Figs. 5(b) and 5(c)]. This synchronization arises from mutual injection locking, mediated by inter-branch crosstalk and resonant enhancement. Mutual injection is facilitated by resonant dispersive waves [DWs, indicated by the dashed black vertical lines shown in panel 1 to 3 in Fig. 5(a)] generated by periodic spatial inhomogeneities of the pulses. In the time domain, DWs appear as wide pedestals with exponentially decaying tails [inset of panel 3 of Fig. 5(b)]. Perturbations from one pulse to adjacent pulses induce attractive or repulsive interactions [7]. During simultaneous pumping, DWs from the two branches interact coherently, generating new DWs and enhancing multimode nonlinear interactions, which in turn mediate pulse interactions and achieve soliton synchronization.

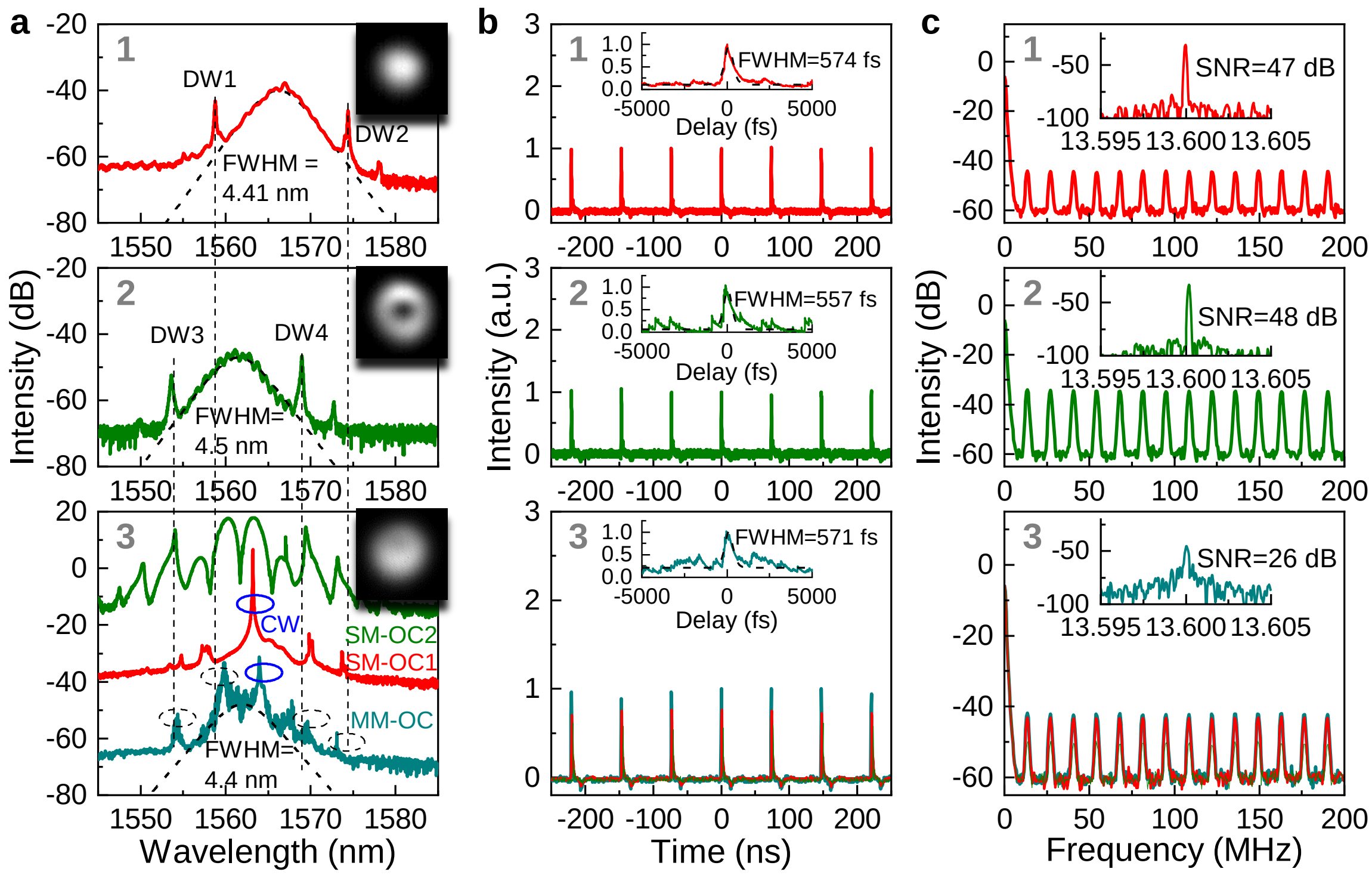

Figure 5: **The characteristics of the spatiotemporal mode-locking in "Trivial" state. (a)**, **(b)** and **(c)** shows respectively the spectra, the temporal train and radio-frequency (RF) spectra for three different working states of 1 to 3. The mode-locking state "1" shown in panels 1 of (a), (b) and (c) are corresponding to the transverse mode $OAM_{00}$ when only pump the branch-1 with the pump power of 51 mW. The inset in panel 1 of (a) is the field distribution of the laser output, the inset in panel 1 of (b) is autocorrelation (AC) curve of the pulse, and the inset in panel 1 of (c) is the RF spectrum with a higher (100 Hz) resolution bandwidth compared to 1 MHz in (c). Similarly, the panels 2 of (a), (b) and (c) and the insets are corresponding to the transverse mode $OAM_{01}$ when only pumping the branch-2 with the pump power of 53 mW. The panels 3 of (a), (b) and (c) and the insets are corresponding to the spatiotemporal mode-locking with $OAM_{00}$ and $OAM_{01}$ when pumping both the branch-1 and the branche-2 with the same settings of the pump power and the PC as in the single pumping in panels 1 and 2.

### 4.2 The spatiotemporal mode-locking in "Topological" state

Figure 6 presents the spatiotemporal mode-locking in "Topological" state, with a time delay of 947.63 ps. Panels 1 in Figs. 6(a) to 6(c) show the optical spectrum, temporal pulse train, and RF spectrum when only branch 1 is pumped at 150 mW. The inset of panel 1 in Fig. 6(a) displays the output field distribution corresponding to the $OAM_{00}$ transverse mode. Panels 2 in Figs. 6(a) to 6(c), together with the inset, correspond to branch 2 ($OAM_{01}$) under the same pump power.

Pumping a single branch alone yields a continuous-wave regime under the given polarization controller (PC) settings: both the pulse train and RF sequence fluctuate randomly, as evident in panel 2 and 3 of Figs. 6(b) and 6(c). When both branches are pumped simultaneously at 150 mW, the system transitions into a stable spatiotemporal mode-locked state, as shown in panel 3 of Figs. 6(a) to 6(c). The MM OC spectrum exhibits two distinct peaks near 1559 nm and 1564 nm, corresponding to mutual injection from branch 1 and branch 2, respectively. Notably, the spectrum under dual-branch pumping is not a simple superposition of the single-branch spectra; the peaks shift slightly relative to their original positions, as indicated by the dashed lines across panel 1 to 3 in Fig. 6(a).

Mode-locked solitons propagate across both single-mode branches at the same longitudinal frequency, as evidenced by overlapping RF spectra measured at SM OC-1, SM OC-2, and MM OC. This demonstrates strong resonant coupling between the branches: the lasing peak of one branch would induce a corresponding peak in the other branch. Such resonant interactions are absent under single-branch pumping, where crosstalk power is largely dissipated by gain-fiber losses in the unpumped branch. The enhanced mutual injection facilitated by resonant coupling is resulted from the periodic potentials induced by the topological lattices, which stabilizes the mode-locking state of the laser system, as discussed latter.

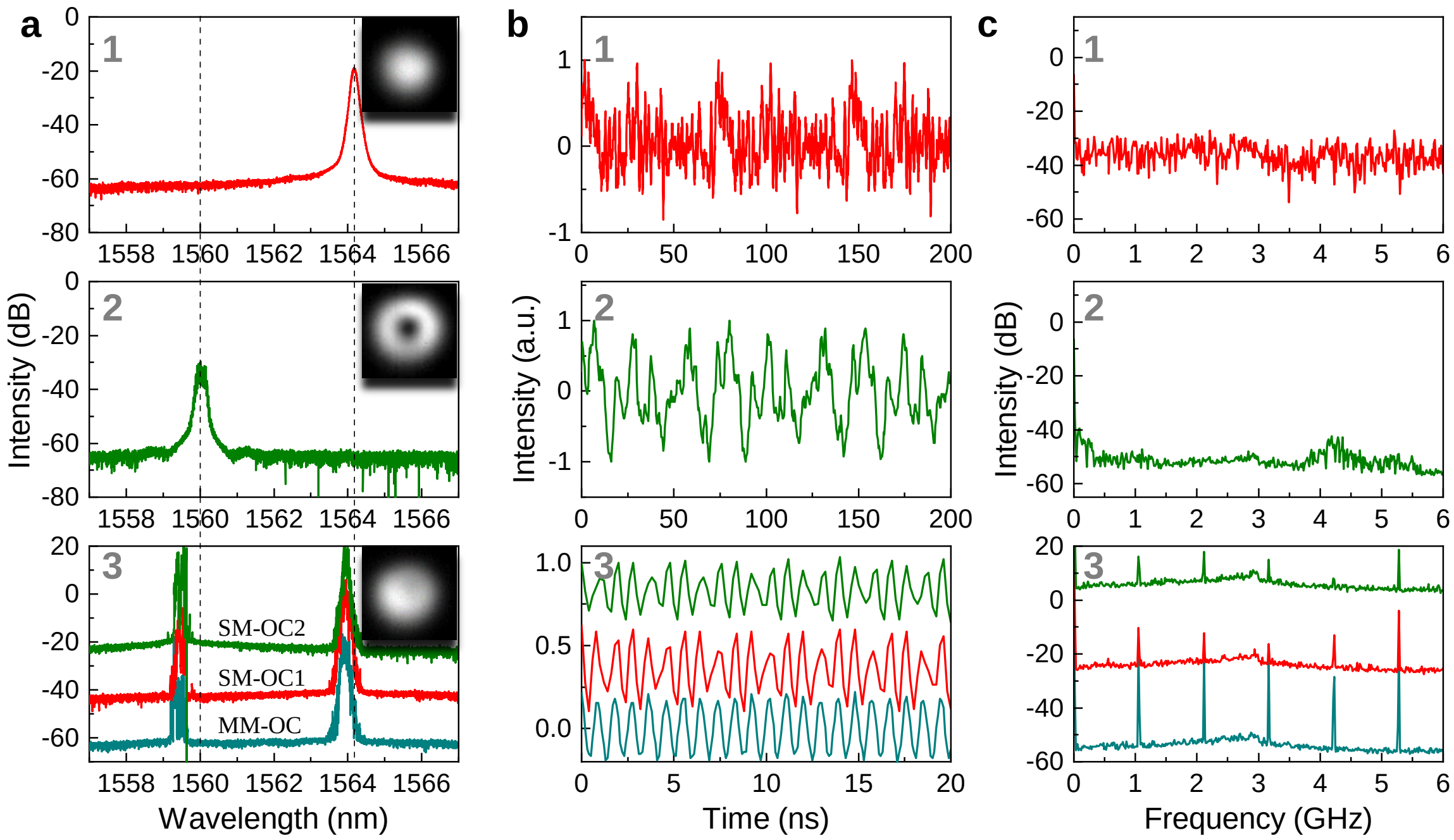

Figure 6: **The characteristics of the spatiotemporal mode-locking in "Topological" state.** (**a**), (**b**) and (**c**) shows respectively the spectra, the temporal train and radio-frequency (RF) spectra for three different working states of 1 to 3. The insets in (a) are the average field distributions of the laser output. The panels 1 in (a), (b) and (c) are corresponding to the transverse mode $OAM_{00}$ when only pump the branch-1 with the pump power of 150 mW. Similarly, the panels 2 in (a), (b) and (c) and the insets are corresponding to the transverse mode $OAM_{01}$ when only pumping the branch-2 with the pump power of 150 mW. The panels 3 of (a), (b) and (c) and the insets are corresponding to the spatiotemporal mode-locking with $OAM_{00}$ and $OAM_{01}$ when pumping both the branch-1 and the branche-2 with the same pump power and the PC settings as the cases of the single pumping in panels 1 and 2.

Figure 7 presents the spectral, beam profile, temporal, and RF characteristics of spatiotemporal mode-locking in the "Topological" state under varying pump powers. In these experiments, branch 1 was fixed at 150 mW, while branch 2 was increased from 150 to 400 mW; all other parameters, including the PC settings and the time delay (947.64 ps), were held constant. The repetition rate of 1.056 GHz remains unchanged across the pump-power range, as evidenced by panel 1 to 3 of Fig. S6B, and the corresponding pulse trains [panel 1 to 3 in Fig. 7(c)] show consistent temporal structure, confirming the pump-power independence of the repetition rate. The optical spectra [panels 1 to 3 of Fig. 7(a)] exhibit a slight blue shift as branch-2 power increases. Simultaneously, the beam profiles in the insets reveal progressively more pronounced $OAM_{01}$ characteristics at higher pump powers. These observations are attributed to enhanced nonlinear effects at higher pump powers, which induce shifts in the phase and consequently the spectral output.

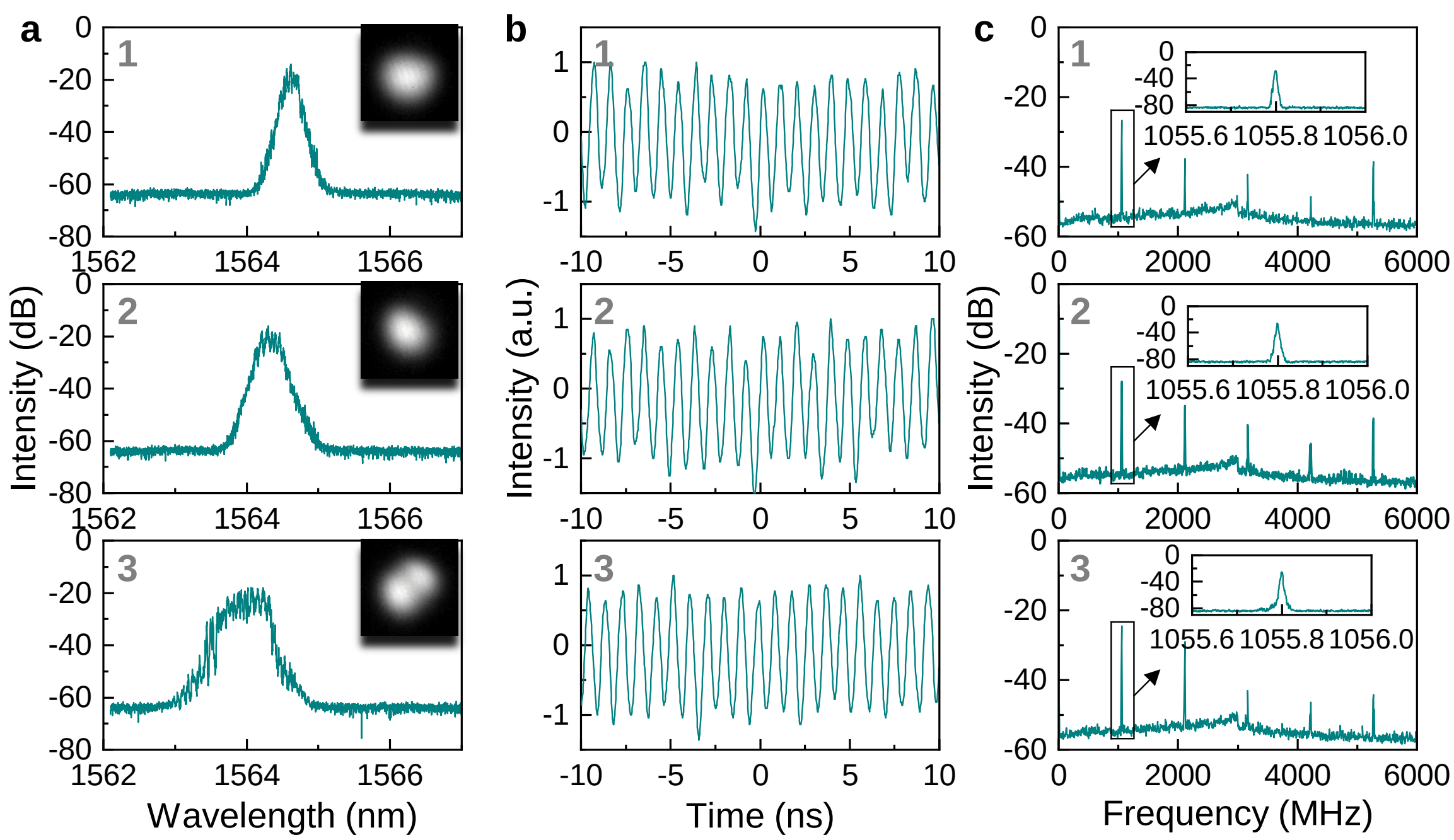

Figure 7: **Spatiotemporal mode-locking with different pumping power.** (**a**), (**b**) and (**c**) shows respectively the spectra, the temporal train and radio-frequency (RF) spectra for three different working states of 1 to 3. The insets in (a) are the field distributions of the laser output. The panels 1 in (a), (b) and (c) are corresponding to the output results of the laser system with pump1 of 150 mW and pump2 of 150 mW. Similarly, the panels 2 in (a), (b) and (c) and the insets are corresponding to to the output results of the laser system with pump1 of 150 mW and pump2 of 300 mW. The panels 3 of (a), (b) and (c) and the insets are corresponding to to the output results of the laser system with pump1 of 150 mW and pump 2 of 400 mW.

### 4.3 Laser power as a function of pump power in different topological states

Figure 8 shows the laser output power as a function of pump power for trivial and topological spatiotemporal mode-locking states. Figs. 8(a) and 8(b) correspond to pumping a single-mode branch individually, whereas Fig. 8(c) shows results for simultaneous pumping of both branches. For the trivial state [Fig. 8(a)], the continuous-wave (CW) lasing thresholds are ~46 mW for $OAM_{00}$ and ~36 mW for $OAM_{01}$, with mode-locking thresholds slightly higher at 48 mW and 45 mW, respectively. The laser power coefficients, extracted from linear fits in the 50–300 mW range, are 0.00553 for $OAM_{00}$ and 0.00530 for $OAM_{01}$.For the topological spatiotemporal mode-locking state [Fig. 8(b)], the CW thresholds are ~46 mW for $OAM_{00}$ and ~38 mW for $OAM_{01}$, with laser power coefficients of 0.00521 and 0.00486, respectively.

When both branches are pumped simultaneously [Fig. 8(c)], the laser power coefficients for trivial and topological spatiotemporal mode-locking are 0.01309 and 0.01137, respectively, which closely approximate the sum of the individual-branch coefficients. This scaling indicates that the output power is effectively the combination of the two single-mode branches. The distinct differences in laser power coefficients highlight the sensitivity of the mode-locked states to the polarization conditions within the cavity.

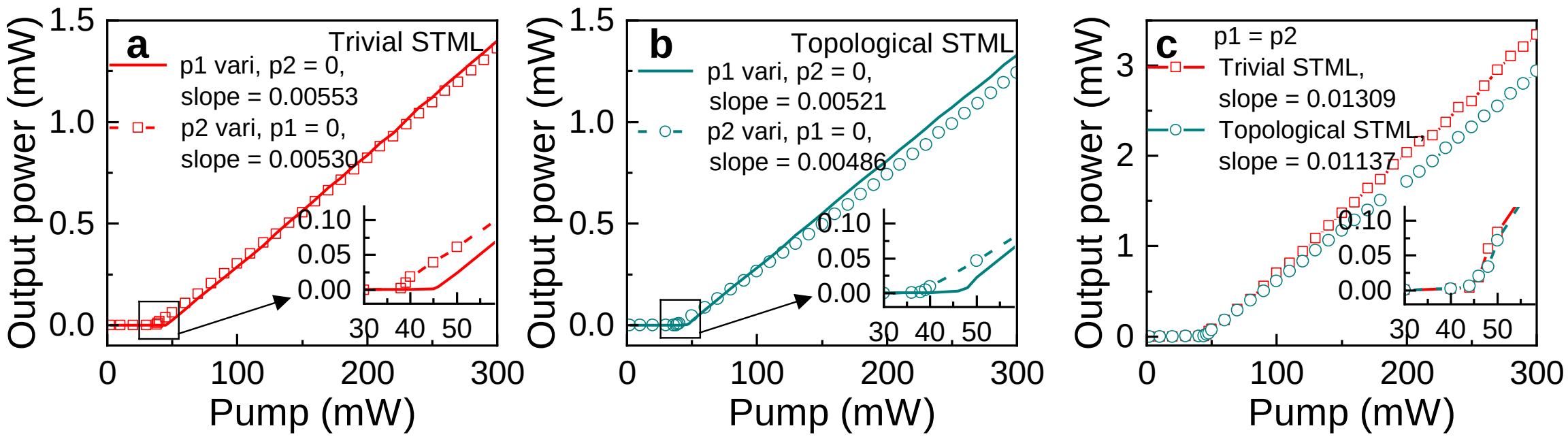

Figure 8: **Laser power as a function of pump power in different topological states.** (**a**) shows the results of pumping only a single-mode branch in the "Trivial" state of $\Delta_t$ = 0 ps for generating the spatiotemporal mode-locking (STML), (**b**) in the "Topological" state of $\Delta_t$ = 947.6 ps for generating the topological STML. (**c**) shows the results of the corresponding topological states when pumping the two single single-mode branches simultaneously.

## 5 Topological protection against cavity noise

In practice, mode-locked fiber lasers are susceptible to environmental noise, such as temperature fluctuations and mechanical vibrations. Environmental perturbations often induce supermode noise or longitudinal mode drifts in mode-locked fiber lasers due to changes in fiber length. In our laser system, instability in the individual delay lines and unintentional path-length mismatches may further distort the delay-induced topological structures, leading to instability in the phase and amplitude of the spatiotemporal pulses. To evaluate the robustness of our topological laser based on MDM configuration, we perform numerical simulations with and without introducing intensity and phase disorders. The obtained results (Fig. 9 and Fig. 10) indicate clear phase instability, soliton drifts and continuous wave (CW) noise in the "Trivial" state, compared with the robustness of the laser against environmental noise in both the "Topological" and "Topological + BEC" states.

Figure 9 and Fig. 10 display the optical spectra, pulse patterns, and phase profiles for the "Topological", "Topological + BEC", and "Trivial" states, respectively, without/with disorders, respectively. For the "Topological" and "Topological + BEC" states, as shown in panels 1 of Figs. 10(a) and 10(b), the pulse patterns exhibit no significant deviation from those in the disorder-free case as shown in Fig. 9. Similarly, the pulse phase, corresponding to a trapping potential around the pulses shown in panels 2 of Figs. 10(a) and 10(b), remains largely unchanged in the central region, with only minor distortions appearing far from the center. The optical spectra shown in panels 3 of Figs. 10(a) and 10(b) also remain consistent, albeit with a slightly elevated noise floor up to approximately –80 dB. These results confirm the robustness of the laser against environmental noise in both the "Topological" and "Topological + BEC" states. In contrast, the laser exhibits clear instability in the "Trivial" state, as evidenced by comparing the disordered results shown in Fig. 10(c) with their disorder-free counterpart in Fig. 9(c). Firstly, the soliton drifts at varying velocities, evidenced by different final positions after 500 round trips, as shown in panel 1 of Fig. S20C. Secondly, the phase profile shown in panel 2 of Fig. 10(c) is severely distorted and partially truncated. Such distortion of the phase potential under noise can undermine the soliton's ability to resist fusion or annihilation due to stochastic interactions and may degrade coherence [18,19]. Thirdly, the noise floor rises to about –50 dB, and a distinct CW peak emerges alongside the pulse spectrum shown in panel 3 of Fig. 10(c), indicating mode-locking instability.

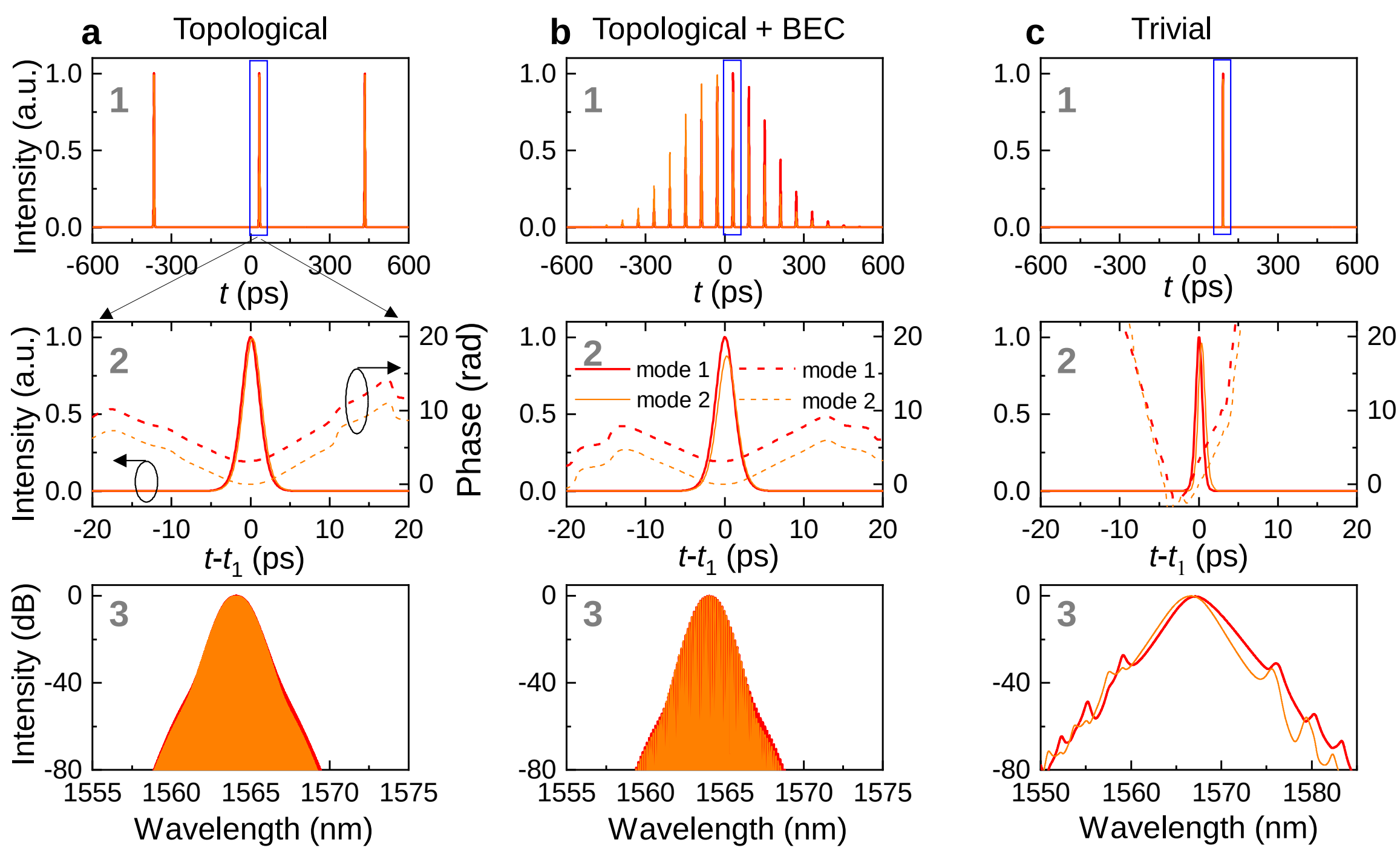

Figure 9: **Pulse patterns and phases in different topological states. (a)-(c)** show respectively the results for the topological states of "Topological", "Topological + BEC" and the "Trivial", simulated without intensity and phase disorders. The panels 1 are the pulse patterns, the panels 2 are the zoomed pulse profile of panels 1 as indicated and their phase curves, and the panels 3 are the corresponding spectra. It should be noted that only the firstly two modes are shown and the other high-order modes are omitted due to low intensity.

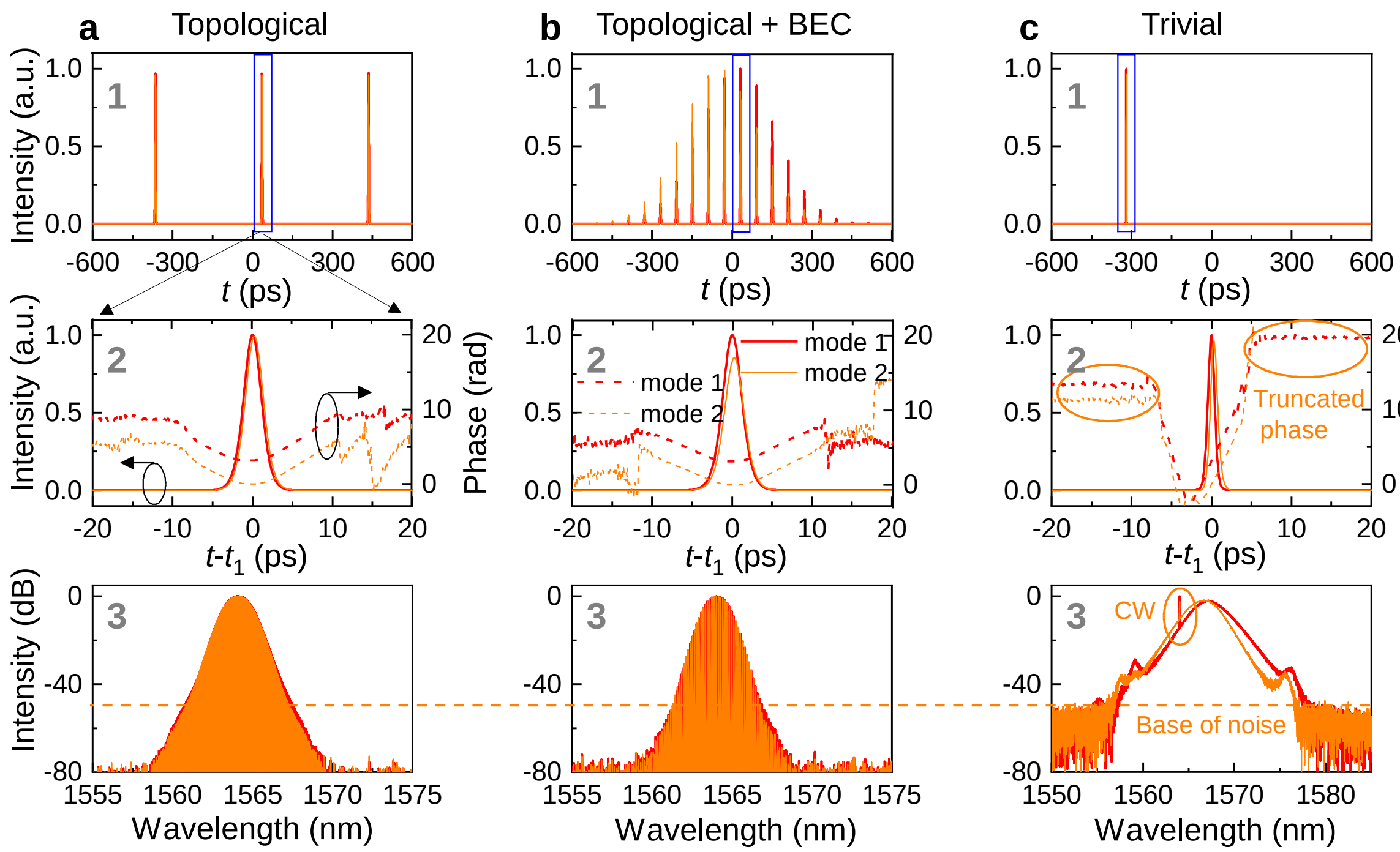

Figure 10: **Simulation results for mode-locking robustness against cavity noise. (a)-(c)** show respectively the results for the topological states of "Topological", "Topological + BEC" and the "Trivial", simulated with intensity and phase disorders of $N(0, 1\text{x}10^{-3}) + iN(0, 1\text{x}10^{-3})$ $W^{-2}$. The panels 1 are pulse patterns, panels 2 are zoomed-in pulse profiles of panels 1 (blue rectangle) overlaid with their phase curves, and panels 3 are corresponding optical spectra. Note that visualizations are limited to the first two modes, and the other high-order modes are excluded due to their low intensity.

The measured RF spectra [Figs. 11(b) and 11(c)] indicate that the mode-locking state remains well confined to the main harmonic frequency peaks given by $f_n = nf = n/\Delta t$ (where $n$ is an integer), with virtually no supermode or spurious noise peaks between them across all three topological states of "Topological", "Topological + BEC", and "Trivial". However, similar to the coexisting CW peaks observed in the simulated optical spectra with intensity and phase noises, measurable CW peaks consistently appear in the optical spectra of the "Trivial" state [panel 1 in Fig. 11(c) and panel 3 in Fig. 10(c)]. In contrast, the "Topological" and "Topological + BEC" states exhibit clean optical spectra without noticeable CW peaks [panel 1 in Figs. 11(a) and 11(b)]. The signal-to-noise ratio (SNR) for the "Trivial" state is considerably lower (~26 dB) compared to that for the "Topological" (~58 dB) and "Topological + BEC" (~55 dB) states. This low SNR indicates that the "Trivial" mode-locking state is more susceptible to environmental perturbations and exbibits relatively poorer stability. In the experiments, it is difficult to sustain mode-locking for a long time in the "trivial" state, unlike the long-term stable mode-locking operation achieved in the "Topological" and "Topological + BEC" states. To enhance mode-locking stability, longitudinal mode drifts and mode beats should be controlled—for instance, by actively stabilizing the cavity length against the perturbations.

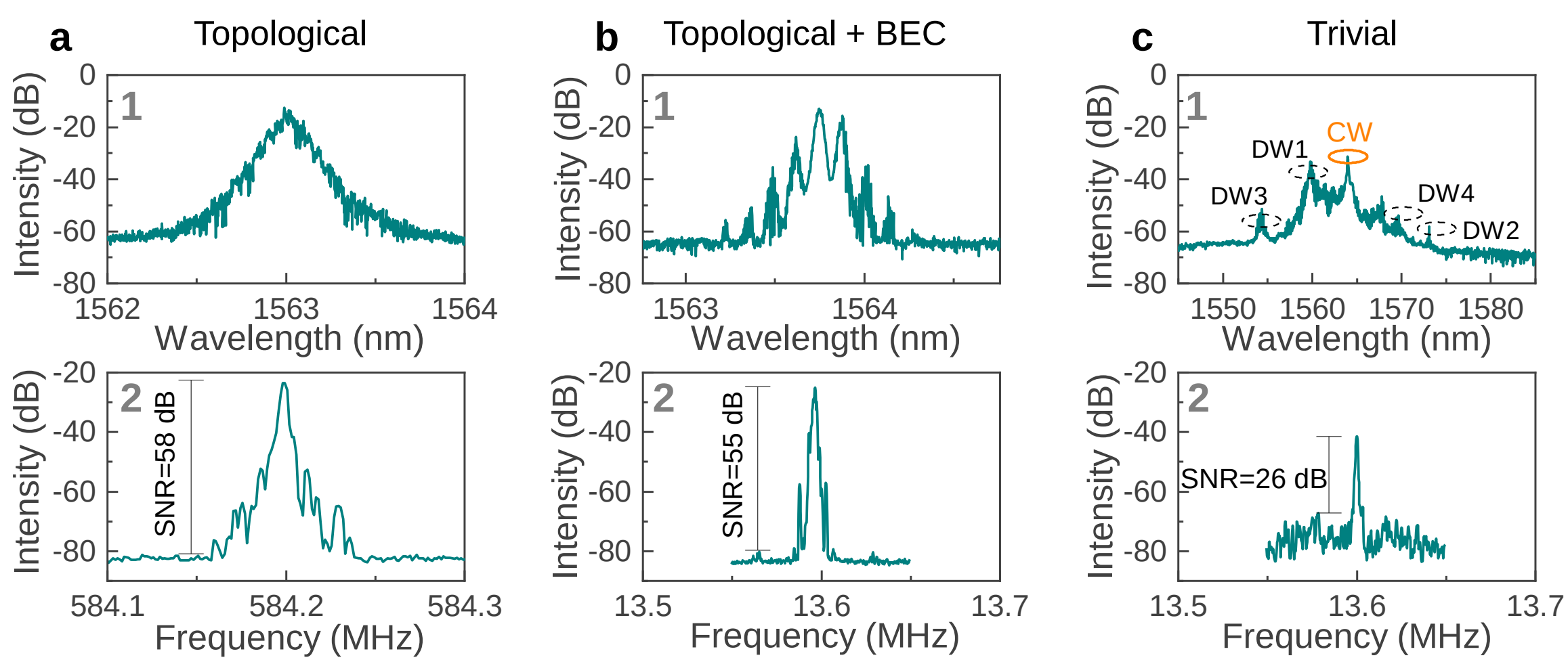

Figure 11: **Experimental results for mode-locking robustness against cavity noise.** (**a**)-(**c**) Characterization of the (a) "Topological", (b) "Topological + BEC", and (c) "Trivial" states. The upper panels 1 display optical spectra, and the lower panels 2 show the corresponding high-resolution RF spectra.

## 6 Additional supplementary graphs

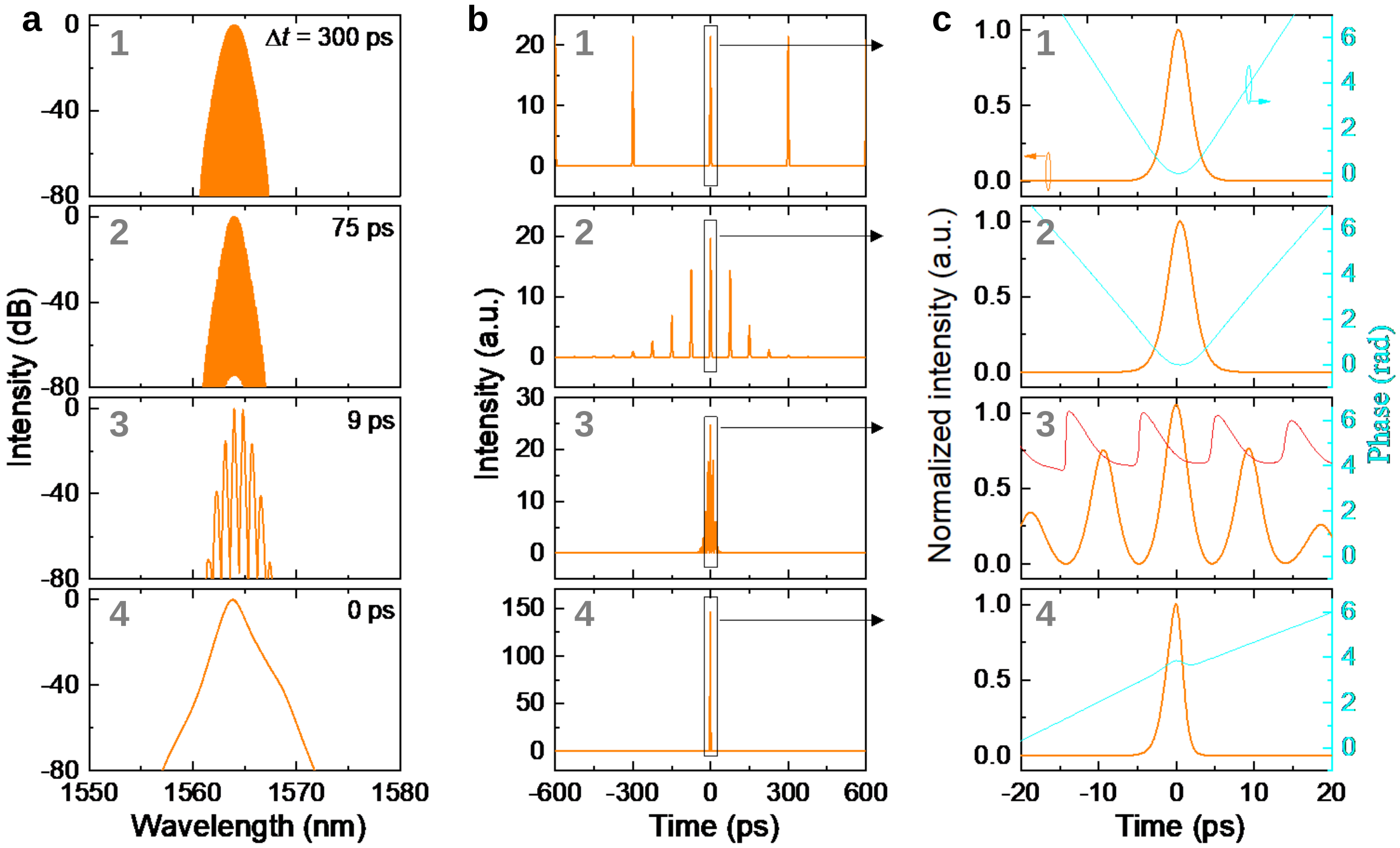

Figure 12: **The soliton condensation as the time delay reduces.** (**a**)-(**c**) show respectively the characteristics of the spectra, pulse train, and the phase for the time delays $\Delta_t$ of 300 ps, 75 ps, 9 ps, 0 ps, corresponding to panels 1 to 4. The simulations were carried out based on the master function with various $\Delta_t$, Esat = 300 pJ and the other parameters are constants. The results were simulated by Eq. (6).

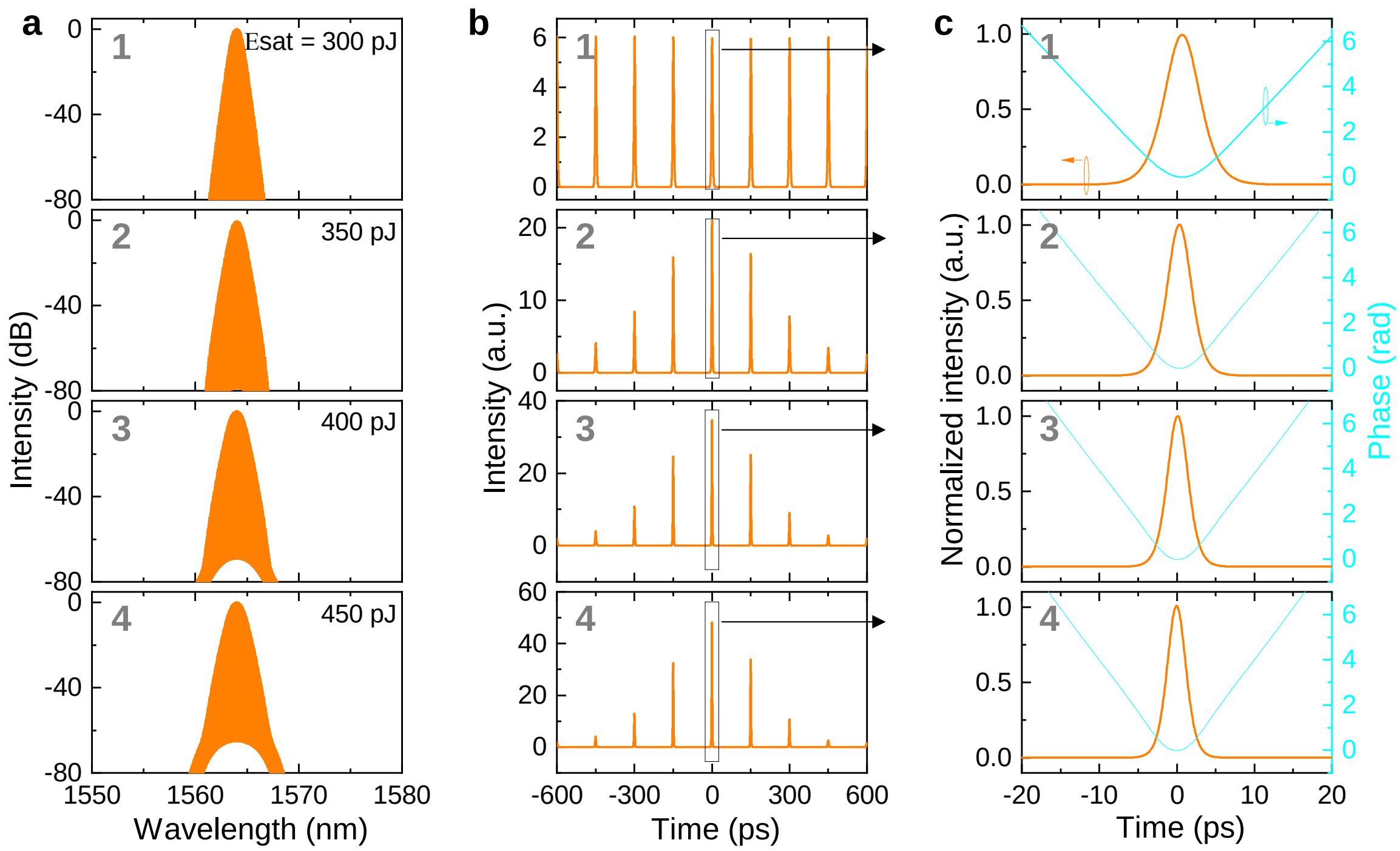

Figure 13: **The soliton condensation as the pump power increases.** (**a**)-(**c**) show respectively the characteristics of the spectra, pulse train, and the phase for the Esat of 300 pJ, 350 pJ, 400 pJ, 450 pJ, corresponding to panels 1 to 4. The simulations were carried out based on the master function with various Esat, $\Delta_t$= 150 ps and the other parameters are constants. Esat is positive correlated with the pump power in the experiments. The results were simulated by Eq. (6).

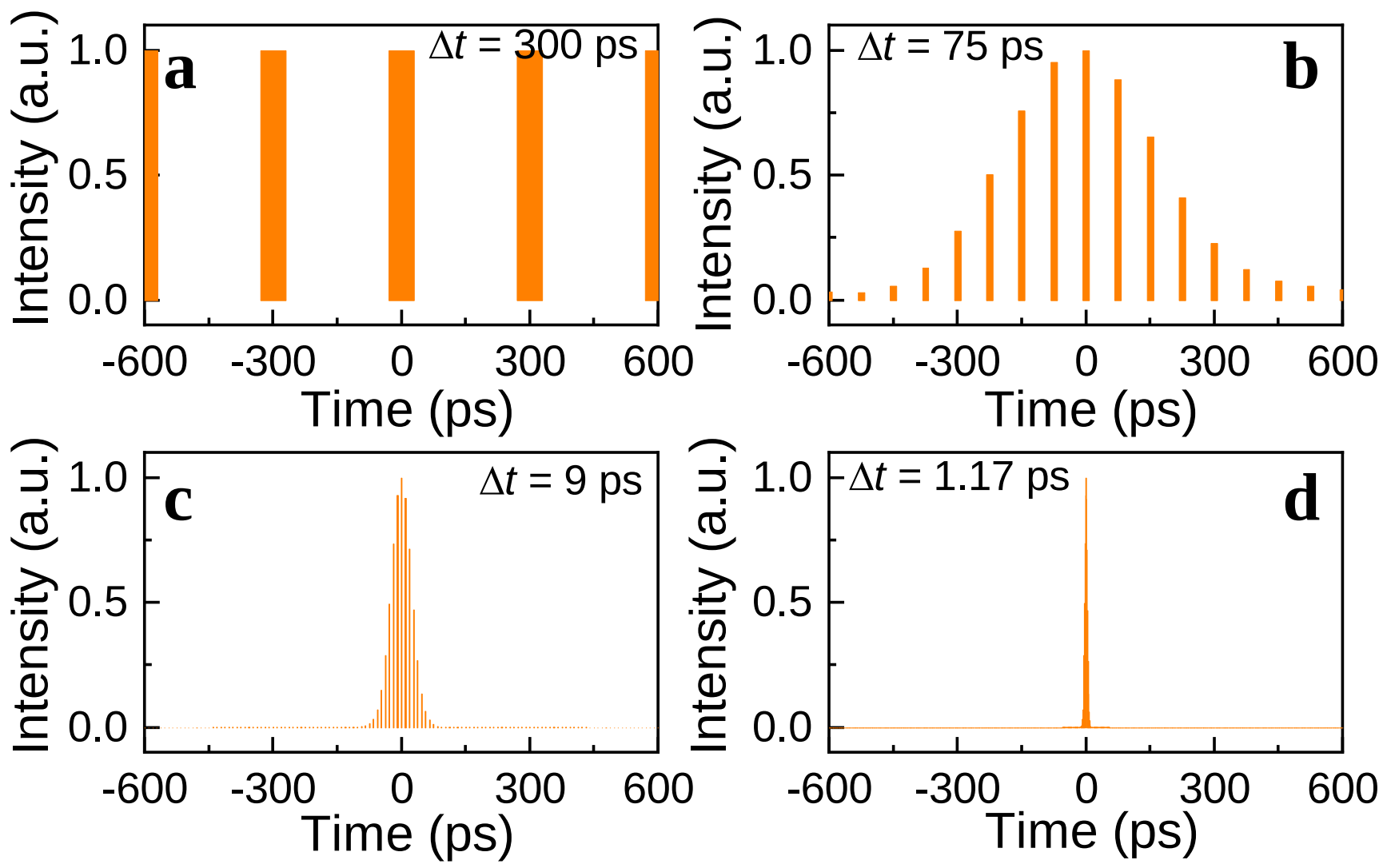

Figure 14: **Simulated results with the tight-binding model.** (**a**)-(**d**) show the pulse train in stable state for the pulse period $\Delta_t$ of 300 ps, 75 ps, 9 ps, 1.17 ps, respectively. The results were simulated by Eq. (11).

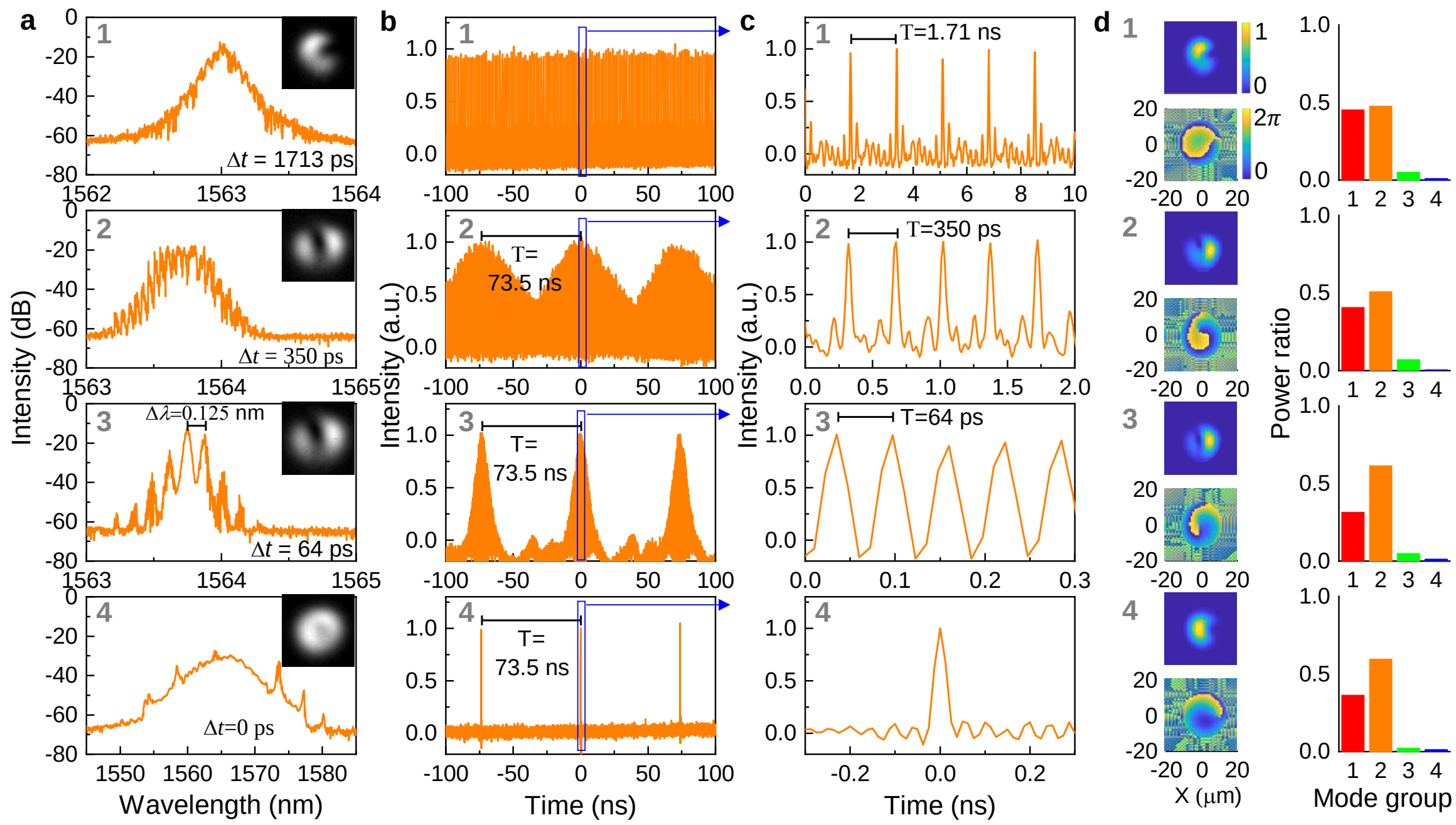

Figure 15: **Measured results of the laser in different spatiotemporal mode-locking states.** (**a**)-(**d**) show the spectrum, the pulse train, the zoomed part of pulse train, and the modal decomposition, respectively. The time-integrated beam profiles are shown in the inset of (**a**). Panels 1-4 show the results for four samples with the time delay $\Delta_t$ of 1713 ps, 350 ps, 64 ps and 0 ps, respectively, corresponding to the results shown in Main Text Fig. 4a.

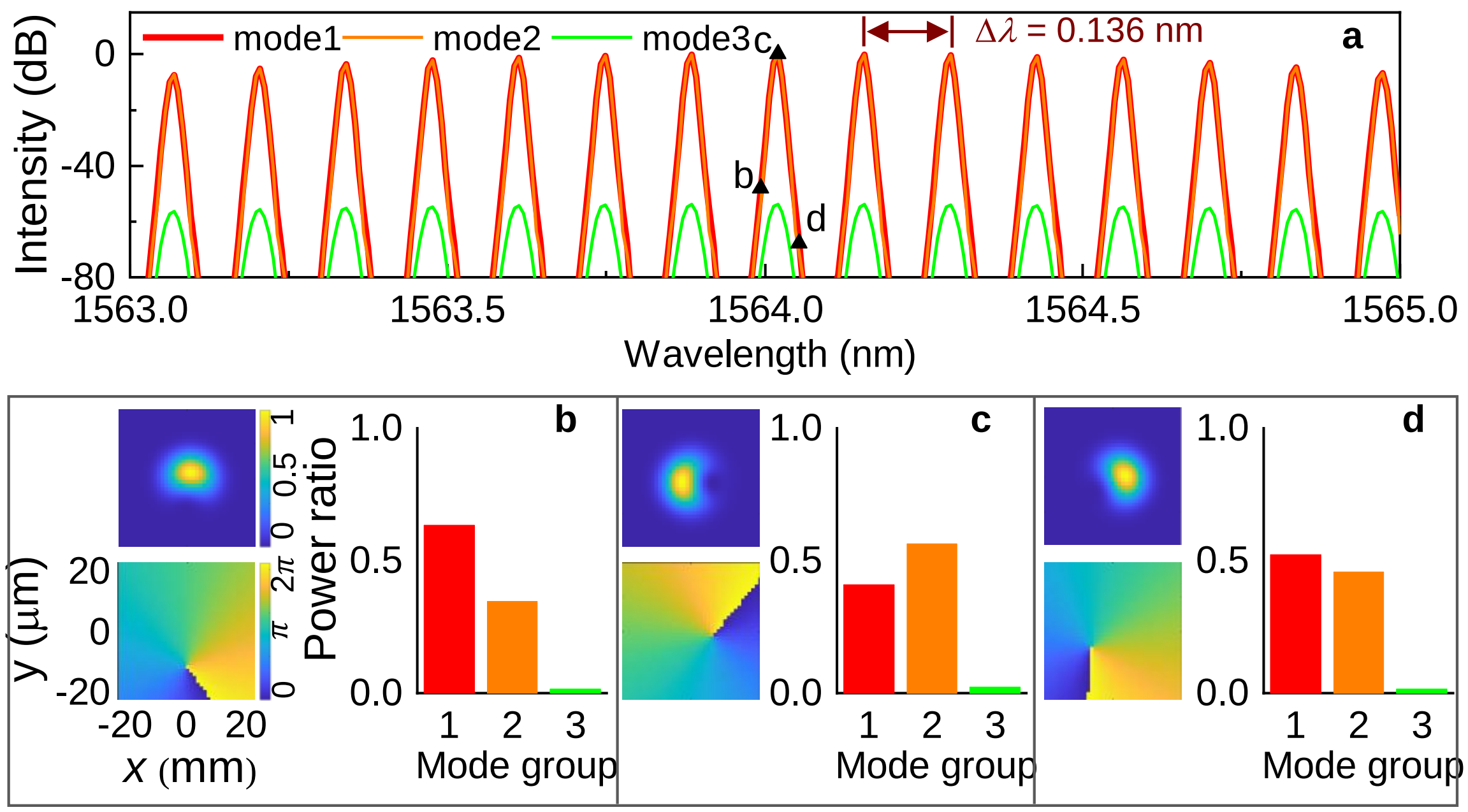

Figure 16: **Spectral resolved modal decomposition and power ratio.** (**a**) A portion of spectrum from the panel 2 in Fig. 2A for $\Delta_t$= 60 ps. (**b**), (**c**) and (**d**) show the spectral resolved beam intensity, phase profiles and the modal decompositions for three spectral positions indicated in (**a**). Mode decompositions at three spectral positions within one modulation period around 1564 nm are shown in (b) to (d). The spectrally resolved beam is dominated by $OAM_{00}$ and $OAM_{01}$, resulting in an intensity distribution resembling an eccentric disk with an off-center singularity, consistent with the phase profile. Both the intensity distribution and the modal power ratio vary with wavelength. The $OAM_{00}$ (mode 1) component is stronger at the spectral positions in (b) and (d), whereas $OAM_{01}$ (mode 2) dominates at the position in (c). Despite these variations, the time-averaged power of $OAM_{00}$ and $OAM_{01}$ remains approximately equal, as shown in Fig. 8, under equal pumping power of the two single-mode branches.

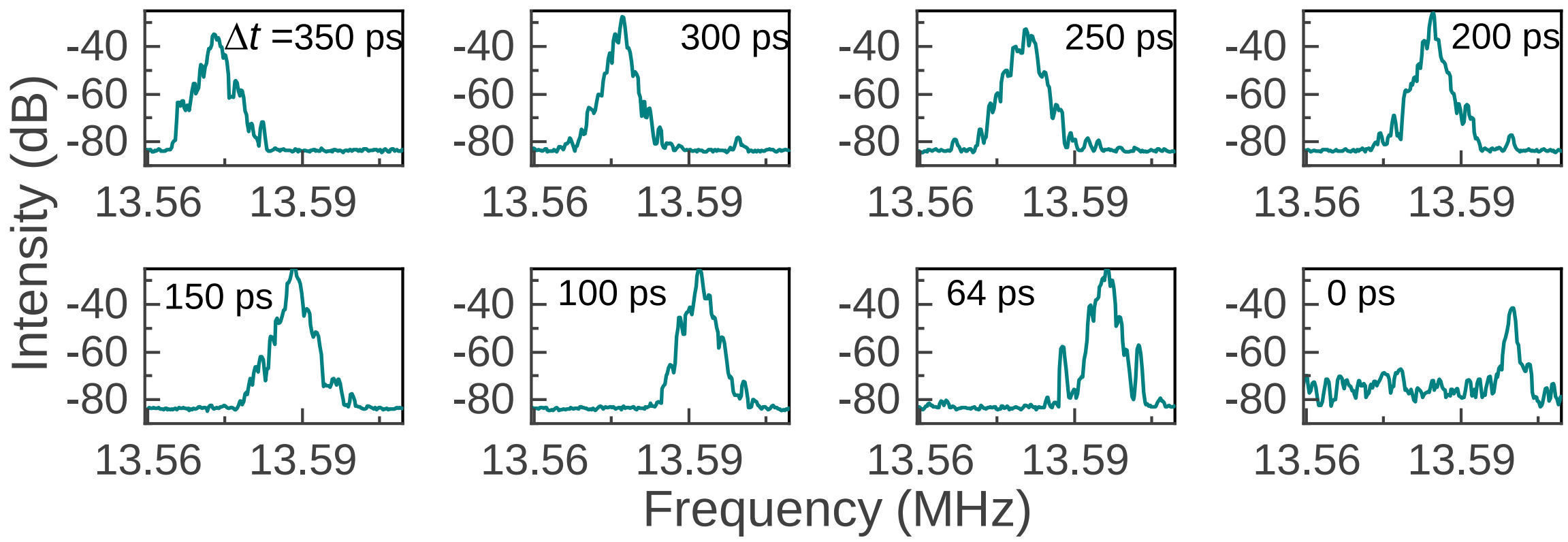

Figure 17: **Evolution of the RF Spectra of fundamental frequency as a function of time delay $\Delta_t$.** The RF spectra measured with a high resolution (bandwidth = 1 kHz) keep almost a similar shape (bandwidth) from $\Delta_t$ = 350 ps to 64 ps. However, the bandwidth of the frequency peak for $\Delta_t$ = 0 ps is obviously reduced, corresponding to the transformation of the topological states.

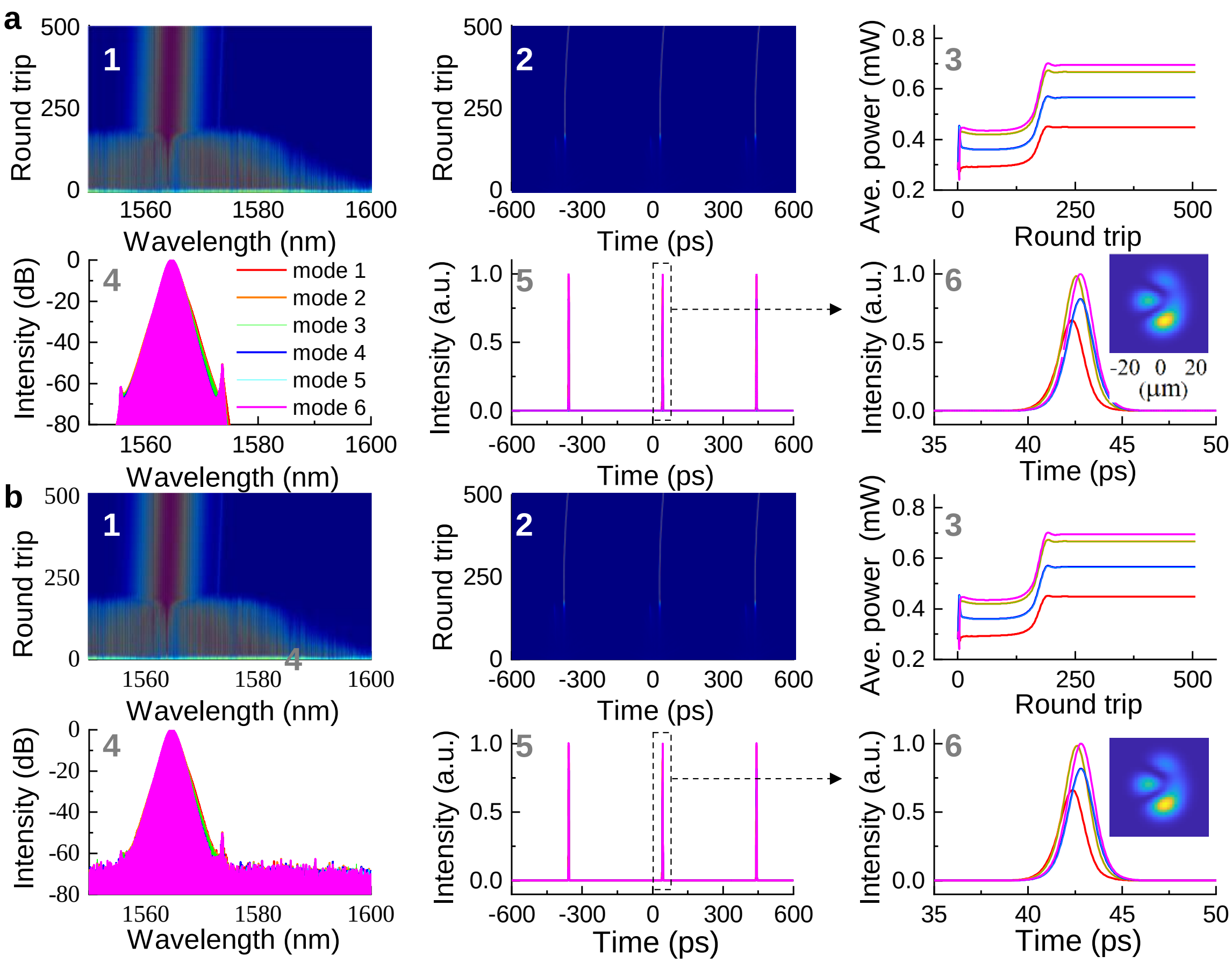

Figure 18: **Simulated results with the lumped model for the spatiotemporal mode-locking in the laser based on the 6-branch MDM.** (**a**) show the results for $\Delta_t$= 400 ps, simulated without intensity and phase disorders. The panels 1 and 4 are for the evolution of the spectra and the spectrum at last round trip, respectively. The panels 2 and 5 are for the evolution of pulse patterns and the pulse train at last round trip, respectively. The panel 3 is for the average modal power as the function of the round-trip. The panel 6 is the zoomed part of pulse from the panel 5, and the inset is the beam profile. Similarly, (**b**) show the results for $\Delta_t$= 400 ps, simulated with intensity and phase disorders of $\mathcal{N}(0,1e\text{-}3) + i\mathcal{N}(0,1e\text{-}3)$ $W^{-2}$.